\documentclass[12pt]{iopart}
\usepackage{graphicx}
\usepackage{bm}
\usepackage[dvips]{color}
\eqnobysec

\newcommand{\beq}{\begin{equation}}
\newcommand{\beqa}{\begin{eqnarray}}
\newcommand{\eeq}{\end{equation}}
\newcommand{\eeqa}{\end{eqnarray}}
\newcommand{\abs}[1]{\left\vert#1\right\vert}
\renewcommand{\d}{{\rm d}}
\newcommand{\de}{{(2)}}
\newcommand{\dpar}{\partial}
\newcommand{\ds}{\displaystyle}
\renewcommand{\e}{{\rm e}}
\newcommand{\eps}{\varepsilon}
\newcommand{\frad}[2]{\ds{\ds#1\over\ds#2}}
\newcommand{\gi}{G_{\rm o}}
\newcommand{\gp}{G_{\rm e}}
\newcommand{\haut}{\rule[-10pt]{0pt}{28pt}}
\newcommand{\hatn}{{\hat n}}

\newcommand{\hatz}{{\hat z}}
\renewcommand{\i}{{\rm i}}
\newcommand{\lbar}{{\overline l}}
\newcommand{\mod}[1]{\ {\rm(mod.}\ #1)}
\renewcommand{\o}{\omega}
\newcommand{\rbar}{{\overline r}}
\newcommand{\si}{\sigma}
\newcommand{\sign}{\mathop{\rm sign}\nolimits}
\newcommand{\so}{_{\rm SO}}
\newcommand{\trois}{{(3)}}
\newcommand{\un}{{(1)}}
\newcommand{\ve}[1]{{\bm#1}}
\renewcommand{\xi}{x_{\rm o}}
\newcommand{\xp}{x_{\rm e}}
\newcommand{\Frac}{\mathop{\rm Frac}\nolimits}
\newcommand{\G}{{\cal G}}
\renewcommand{\H}{{\cal H}}
\newcommand{\Int}{\mathop{\rm Int}\nolimits}
\renewcommand{\O}{{\cal O}}
\newcommand{\PT}{{\mathrm P}}

\begin{document}

\title{Magnetization of two coupled rings}

\author{Y Avishai$^{1,2}$ and J M Luck$^3$}

\address{$^1$ Department of Physics and Ilse Katz Center for Nanotechnology,
Ben Gurion University, Beer Sheva 84105, Israel}

\address{$^2$ Department of Physics, Hong Kong University
of Science and Technology, Clear Water Bay, Kowloon, Hong Kong}

\address{$^3$ Institut de Physique Th\'eorique,
IPhT, CEA Saclay, and URA 2306, CNRS, 91191~Gif-sur-Yvette cedex, France}

\begin{abstract}
We investigate the persistent currents and magnetization
of a mesoscopic system consisting of two clean metallic rings
sharing a single contact point in a magnetic field.
Many novel features with respect to the single-ring geometry are underlined,
including the explicit dependence of wavefunctions on the Aharonov-Bohm fluxes,
the complex pattern of twofold and threefold degeneracies,
the key r\^ole of length and flux commensurability,
and in the case of commensurate ring lengths
the occurrence of idle levels which do not carry any current.
Spin-orbit interactions,
induced by the electric fields of charged wires threading the rings,
give rise to a peculiar version of the Aharonov-Casher effect where,
unlike for a single ring, spin is not conserved.
Remarkably enough, this can only be realized when the Aharonov-Bohm fluxes
in both rings are neither integer nor half-integer multiples
of the flux quantum.
\end{abstract}


\eads{\mailto{yshai@bgu.ac.il},\mailto{jean-marc.luck@cea.fr}}

\maketitle

\section{Introduction}

The orbital magnetic response of mesoscopic systems has been extensively
studied in the last 25 years~\cite{Imry}.
One of its manifestations is the occurrence of persistent currents
and weak magnetism in small (sub-micron size) metallic rings threaded
by a magnetic flux~\cite{BIL}.
By `small' one means that the circumference $L$
of the ring is much smaller than the phase coherence length $L_\phi$,
and so quantum coherence is maintained throughout.
This condition can be achieved at low enough temperatures.
In a metallic ring disorder is very weak
(as expressed in terms of the Ioffe-Regel condition $k_F\ell\gg 1$,
with $k_F$ and~$\ell$ being the Fermi momentum and the mean free path),
and the currents persist for systems of a few microns in size~\cite{PCEXP},
while for $k_F\ell\le 1$ they decay exponentially with the system size.
The magnetic response of metallic rings proved to be an important tool
to study fundamental aspects of quantum mechanics,
such as quantum coherence~\cite{MH}
and the Aharonov-Bohm (AB) effect~\cite{AB,ABEXP}.
In systems exhibiting the AB effect,
the magnetic flux appears as a pure Abelian U(1) gauge,
in the sense that it only affects the phase of the wavefunction.
Every observable quantity is therefore periodic in the
magnetic flux (this is the Byers-Yang theorem~\cite{BY}).
Analogously,
in the presence of a strong electric field
(either internal or external),
spin-orbit interactions give rise to
the Aharonov-Casher (AC) effect~\cite{AC,MS},
which can be described as a pure non-Abelian~SU(2)~gauge.

\begin{figure}[!ht]
\begin{center}
\includegraphics[angle=90,width=.4\linewidth]{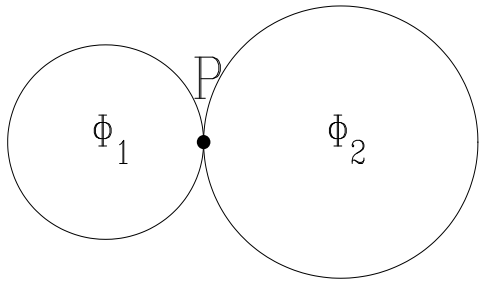}
\caption{The sample considered in this work consists of two clean wires
modeled as ideal rings
with lengths $L_1$ and $L_2$, areas $A_1$ and $A_2$,
threaded by fluxes $\Phi_1$ and $\Phi_2$,
touching at a contact point~P.}
\label{figtwo}
\end{center}
\end{figure}

In this work, motivated by these issues, we examine
the relevance of the topology of the sample to its magnetic response.
We investigate thoroughly a closed system composed of two (ideal)
metallic rings sharing a single contact point P (see Figure~\ref{figtwo}).
From a topological viewpoint,
this sample has genus two, whereas a single ring has genus one.
This difference turns out to enrich the physics in a rather non-trivial way,
as already noticed in~\cite{Schmeltzer}.
The present work includes a systematic study of the energy spectrum,
persistent currents and magnetization of the sample
shown in Figure~\ref{figtwo},
including their dependence on the ring lengths,
on the magnetic fluxes,
and on the number $N$ of electrons at zero temperature.
The difference between the single and double-ring geometry
is underlined throughout.
The main novel features of the latter case are as follows.
Wavefunctions generically depend on the fluxes.
A rich pattern of twofold and threefold degeneracies is observed.
As a consequence, the question whether a given level
has a paramagnetic or diamagnetic response
is more subtle than in the single-ring geometry.
The commensurability of both ring lengths is a key issue.
In the case of commensurate ring lengths,
a finite fraction of the levels are `idle',
in the sense that their energies do not depend on the magnetic fluxes,
so that these levels do not contribute to the persistent currents.
The case of incommensurate ring lengths requires a special treatment,
inspired from the theory of modulated incommensurate structures.
As for the AC effect, the two-ring geometry allows one to study a novel feature
which is absent in the single-ring geometry.
Suppose that the AC effect is realized by threading a ring
with a long straight wire with constant longitudinal charge density.
In a single-ring geometry this construction implies that $s_z$,
the spin component along the ring axis, is conserved,
and so the problem decomposes into two independent ones,
one for spin up and one for spin down.
In the two-ring geometry,
we can realize the AC effect as a pure gauge in an~$s_z$ non-conserving system.
Remarkably enough,
this is possible only if the magnetic fluxes through both rings
are non-trivial, i.e., neither integer nor half-integer multiples
of the flux quantum.
This kind of interplay between the AB and AC effects
had not been noticed so far, to the best of our knowledge.
Finally, the introduction of a non-Abelian SU(2) flux
can also affect the sign of the sample's magnetization,
turning a diamagnetic to a paramagnetic response or {\it vice versa}.

In the following we elaborate and substantiate the issues presented above.
Star\-ting with the AB effect, the following topics are successively covered
(section num\-bers in parentheses):
the Hamiltonian and its characteristic equation~(\ref{secham}),
basic observables, including persistent currents and magnetization~(\ref{PCM}),
various special cases of interest~(\ref{special}),
twofold and threefold degeneracies~(\ref{degene}),
and the spectrum and observables for commensurate~(\ref{com})
and incommensurate~(\ref{incom}) ring lengths.
Then in Section~\ref{so} we examine the r\^ole of spin-orbit interactions
and construct a Hamiltonian in terms of SU(2) fluxes leading to the AC effect.
The energy levels and the magnetization are calculated
in the special case of two equal rings,
the emphasis being put on a novel AB-AC interference effect.
A summary of our findings and a discussion are presented
in Section~\ref{Sec:Discussion},
while two appendices are devoted to a reminder of the well-known case
of a single ring~(A),
and to an extension of the analysis to three coupled rings~(B).

\section{The Hamiltonian and its characteristic equation}
\label{secham}

We consider a clean metallic sample in the form of two unequal rings
touching at a contact point P, as shown in Figure~\ref{figtwo}.
The rings are planar, but may otherwise assume arbitrary shapes.
The rings have lengths $L_1$ and $L_2$ and areas $A_1$ and~$A_2$.
In the presence of a uniform transverse magnetic field $B$,
they are therefore threaded by magnetic fluxes $\Phi_1=BA_1$ and $\Phi_2=BA_2$.
In the case of circular rings,
to be used in numerical illustrations of our results,
we have $A_1=L_1^2/(4\pi)$ and $A_2=L_2^2/(4\pi)$.
In order to compare both ring lengths,
we introduce a variable $0<\o<1$ so that
\beq
\frac{L_1}{L_1+L_2}=\o,\quad\frac{L_2}{L_1+L_2}=1-\o.
\label{odef}
\eeq
It will turn out that the system has different characteristics
for commensurate lengths (rational $\o$)
and for incommensurate lengths (irrational $\o$).

To write down the Hamiltonian, it is useful to employ reduced units
($\hbar=c=e=2m=1$),
so that the flux quantum reads $\Phi_0=2\pi$.
The spectrum and the observables are therefore $2\pi$-periodic
in $\Phi_1$ and $\Phi_2$.
We parametrize a point of the left ring
by its curvilinear abscissa $0\le s_1\le L_1$,
starting from the contact point P and oriented clockwise,
and similarly a point of the right ring by $0\le s_2\le L_2$.
Furthermore, we neglect spin degrees of freedom (except in Section~\ref{so}).
The one-body Hamiltonian of the system reads
\beq
\H=(p_1-a_1)^2+(p_2-a_2)^2,
\eeq
with
\beq
p_1=-\i\,\frac{\d}{\d s_1},\quad
p_2=-\i\,\frac{\d}{\d s_2},
\label{pdef}
\eeq
whereas the tangential vector potentials $a_1$ and $a_2$ can be taken equal to
\beq
a_1=\frac{\Phi_1}{L_1},\quad a_2=\frac{\Phi_2}{L_2}.
\label{adef}
\eeq
A state $\vert\psi\rangle$
is described by a pair of wavefunctions $\{\psi^\un(s_1),\;\psi^\de(s_2)\}$.
The first term of the Hamiltonian $\H$
acts on the left component $\psi^\un(s_1)$,
whereas the second term acts on the right component $\psi^\de(s_2)$.
The behavior of the wavefunctions at the contact point~P
is in general described by a unitary junction $S$-matrix.
In the present work we make the simplest choice, which corresponds
to just requiring the continuity of the wavefunction:
\beq
\psi^\un(0)=\psi^\un(L_1)=\psi^\de(0)=\psi^\de(L_2)=\psi(\PT),
\label{pcont}
\eeq
and the conservation of the current:
\beqa
&&(p_1-a_1)\psi^\un(0)-(p_1-a_1)\psi^\un(L_1)\cr
&&{\hskip -14.5pt}+(p_2-a_2)\psi^\de(0)-(p_2-a_2)\psi^\de(L_2)=0.
\label{pcur}
\eeqa

Setting $E=q^2$ for the energy eigenvalue,
we look for an eigenstate of $\H$ in the form
\beq
\matrix{
\psi^\un(s_1)=\e^{\i a_1s_1}(A_1\e^{\i qs_1}+B_1\e^{-\i qs_1}),\hfill\cr
\psi^\de(s_2)=\e^{\i a_2s_2}(A_2\e^{\i qs_2}+B_2\e^{-\i qs_2}).\hfill}
\label{wform}
\eeq
The condition~(\ref{pcont}) allows one to express
these four amplitudes in terms of~$\psi(\PT)$~as
\beq
\matrix{
A_1=\frad{\e^{-\i\Phi_1}-\e^{-\i qL_1}}{2\i\sin qL_1}\,\psi(\PT),\hfill&
B_1=\frad{\e^{\i qL_1}-\e^{-\i\Phi_1}}{2\i\sin qL_1}\,\psi(\PT),\hfill\cr
A_2=\frad{\e^{-\i\Phi_2}-\e^{-\i qL_2}}{2\i\sin qL_2}\,\psi(\PT),\hfill&
B_2=\frad{\e^{\i qL_2}-\e^{-\i\Phi_2}}{2\i\sin qL_2}\,\psi(\PT).\hfill}
\label{abcdres}
\eeq
The condition~(\ref{pcur}) then yields the characteristic equation
\beq
D(q)=0,
\label{dzero}
\eeq
where the characteristic function reads
\beq
D(q)=\sin q(L_1+L_2)-\cos\Phi_2\,\sin qL_1-\cos\Phi_1\,\sin qL_2.
\label{dres}
\eeq

The eigenstates of the Hamiltonian correspond to
the solutions~$q_n\ge0$ of~(\ref{dzero}), labeled by $n=1,2,\dots$
and ordered as $0\le q_1\le q_2\le\dots$
The corresponding energy eigenvalues are $E_n=q_n^2$.
In the special case where each ring is threaded by a quarter flux unit
($\Phi_1=\Phi_2=\pi/2$),
we have $q_n=n\pi/(L_1+L_2)$~(see~(\ref{linderiv})).
In the general case, i.e., for arbitrary values of the fluxes, we~set
\beq
q_n=\frac{n\pi+g_n}{L_1+L_2}\quad(n=1,2,\dots).
\label{gdef}
\eeq
We will refer to $g_n$ as the {\it modulation} of the spectrum of eigenmomenta
with respect to the linear behavior~(\ref{linderiv}).
This quantity will be shown in Section~\ref{incom} to obey the bound
\beq
\abs{g_n}\le\pi.
\label{gbound}
\eeq

The normalization of the wavefunction of the $n$th level reads
\beq
\langle\psi_n\vert\psi_n\rangle
=\int_0^{L_1}\abs{\psi_n^\un(s_1)}^2\d s_1
+\int_0^{L_2}\abs{\psi_n^\de(s_2)}^2\d s_2
=\abs{\psi_n(\PT)}^2\Delta(q_n),
\eeq
where
\beq
\Delta(q_n)=L_1\,\frac{1-\cos\Phi_1\,\cos q_nL_1}{\sin^2 q_nL_1}
+L_2\,\frac{1-\cos\Phi_2\,\cos q_nL_2}{\sin^2 q_nL_2},
\label{nres}
\eeq
hence
\beq
\psi_n(\PT)=\Delta(q_n)^{-1/2},
\label{norres}
\eeq
up to an irrelevant phase factor.

It is worth underlining a key difference between
the present situation and that of a single ring, recalled in Appendix~A.
The gauge transformation employed in~(\ref{wform}),
while it locally eliminates the vector potential from the
Schr\"odinger equation,
does not lead to a Bloch-type boundary condition,
at variance with~(\ref{Eq_gauge}).
As a consequence, and in contrast with the single-ring wavefunction $\psi(s)$,
the functions $\psi_1(s)$ and $\psi_2(s)$ given in~(\ref{wform})
are not periodic in their respective arguments $s_1$ and $s_2$.
The two-ring topology is indeed in marked difference with the single-ring one.
Waves propagating in each ring are scattered
at each passage at the contact point~P.
This multiple-scattering phenomenon destroys the periodicity
of the plane waves characteristic of the single-ring problem.
As a consequence, stationary states, as given by solutions of~(\ref{dzero}),
bear in general a non-trivial dependence on both ring lengths $L_1, L_2$
and on both fluxes $\Phi_1,\Phi_2$.
The periodicity of physical observables in the fluxes
is however guaranteed by the Byers-Yang theorem,
whose validity is independent of whether there is a Bloch analogue or~not.

\section{Persistent currents and magnetization}
\label{PCM}

The contributions of the $n$th level to the persistent currents
in each ring and to the magnetization read
\beq
I_{1,n}=-\frad{\dpar E_n}{\dpar\Phi_1},\quad
I_{2,n}=-\frad{\dpar E_n}{\dpar\Phi_2},\quad
M_n=-\frad{\dpar E_n}{\dpar B}=A_1I_{1,n}+A_2I_{2,n}.
\label{imdef}
\eeq
The persistent currents can be evaluated as follows.
Considering $I_{1,n}$ for definiteness, we have
\beqa
I_{1,n}
&=&-\frac{1}{L_1}\,\frac{\dpar E_n}{\dpar a_1}
=\frac{2}{L_1}\,\langle\psi_n\vert p_1-a_1\vert\psi_n\rangle\cr
&=&\frac{2}{L_1}\,\int_0^{L_1}\psi_n^{\un\star}(s_1)
\left(-\i\,\frac{\d}{\d s_1}-a_1\right)\psi_n^\un(s_1)\,\d s_1.
\label{iex}
\eeqa
Using the expression~(\ref{wform}) of the normalized wavefunction,
together with~(\ref{abcdres}), (\ref{nres}) and~(\ref{norres}),
we obtain after some algebra
\beq
I_{1,n}=\frac{2q_n}{L_1}\,Q_{1,n},\quad
I_{2,n}=\frac{2q_n}{L_2}\,Q_{2,n},\quad
M_n=2q_n\left(\frac{A_1Q_{1,n}}{L_1}+\frac{A_2Q_{2,n}}{L_2}\right),
\label{im}
\eeq
where the dimensionless current amplitudes read
\beq
Q_{1,n}=-\frac{L_1\,\sin\Phi_1}{\sin q_nL_1\,\Delta(q_n)},\quad
Q_{2,n}=-\frac{L_2\,\sin\Phi_2}{\sin q_nL_2\,\Delta(q_n)}.
\label{qres}
\eeq
An alternative approach consists in evaluating the currents from the spectrum.
We have
\beq
I_{1,n}=-2q_n\,\frac{\dpar q_n}{\dpar\Phi_1}
=2q_n\left(\frac{\dpar D/\dpar\Phi_1}{\dpar D/\dpar q}\right)_{q=q_n}
=2q_n\,\frac{\sin\Phi_1\,\sin q_nL_2}{(\dpar D/\dpar q)_{q=q_n}}.
\label{iex2}
\eeq
The current amplitudes can therefore be expressed as
\beq
Q_{1,n}=-L_1\,\frac{\dpar q_n}{\dpar\Phi_1}
=-\o\,\frac{\dpar g_n}{\dpar\Phi_1},\quad
Q_{2,n}=-L_2\,\frac{\dpar q_n}{\dpar\Phi_2}
=-(1-\o)\,\frac{\dpar g_n}{\dpar\Phi_2}.
\label{qalt}
\eeq
The following identity ensures that the results~(\ref{im}),~(\ref{qres})
and~(\ref{iex2}),~(\ref{qalt}) are identical:
\beq
\left(\frac{\dpar D}{\dpar q}\right)_{q=q_n}
=-\sin q_nL_1\,\sin q_nL_2\,\Delta(q_n).
\label{derex}
\eeq

For a zero-temperature system with $N$ electrons,
the $N$ lowest energy states are occupied.
The total energy and magnetization are therefore given by
\beq
E=\sum_{n=1}^NE_n,\quad M=-\frac{\dpar E}{\dpar B}=\sum_{n=1}^NM_n.
\label{total}
\eeq

Finally, explicit bounds on various quantities of interest
can be derived by applying the inequality
$\abs{\langle\psi_n\vert\O\vert\psi_n\rangle}^2\le
\langle\psi_n\vert\O^2\vert\psi_n\rangle$
to the operators $\O_1=p_1-a_1$ and $\O_2=p_2-a_2$.
The expression~(\ref{iex}) implies that the persistent currents obey the bound
\beq
(L_1\,I_{1,n})^2+(L_2\,I_{2,n})^2\le4q_n^2,
\label{ibound}
\eeq
i.e.,
\beq
Q_{1,n}^2+Q_{2,n}^2\le1.
\label{qbound}
\eeq
Furthermore, using~(\ref{iex2}) and~(\ref{derex}),
we can respectively recast~(\ref{ibound}) as
\beq
\Delta(q_n)^2\ge
\left(\frac{L_1\sin\Phi_1}{\sin q_nL_1}\right)^2+
\left(\frac{L_2\sin\Phi_2}{\sin q_nL_2}\right)^2
\eeq
and
\beq
\left(\frac{\dpar D}{\dpar q}\right)^2_{q=q_n}\ge
(L_1\sin\Phi_1\,\sin q_nL_2)^2+(L_2\sin\Phi_2\,\sin q_nL_1)^2.
\label{derbound}
\eeq

\section{Special cases of interest}
\label{special}

\subsection{No magnetic fluxes}
\label{absence}

We consider first the problem in the absence of fluxes:
$\Phi_1=\Phi_2=0\mod{2\pi}$.\footnote{We recall that
the notation$\mod{2\pi}$ means {\it up to a multiple of} $2\pi$.}
The characteristic function~(\ref{dres}) factors as
\beq
D(q)=-4\,\sin\frac{q(L_1+L_2)}{2}\,\sin\frac{qL_1}{2}\,\sin\frac{qL_2}{2}.
\label{dfactor}
\eeq
The spectrum therefore consists of the following three sectors,
in correspondence with the factors of the above expression.

\begin{itemize}

\item
{\it Bilateral states}.
These states correspond to $\sin q(L_1+L_2)/2=0$, hence
\beq
q=\frac{b\pi}{L_1+L_2}\quad(b=0,2,4,\dots),
\label{q1}
\eeq
with `$b$' for bilateral.
The corresponding wavefunctions are standing waves living on the whole system:
\beq
\matrix{
\psi^\un(s_1)=\psi(\PT)\,\frad{\cos qs_1+\cos q(L_1-s_1)}{1+\cos
qL_1},\hfill\cr
\psi^\de(s_2)=\psi(\PT)\,\frad{\cos qs_2+\cos q(L_2-s_2)}{1+\cos qL_1}.\hfill}
\eeq

\item
{\it Left states}.
These states correspond to $\sin qL_1/2=0$, hence
\beq
q=\frac{l\pi}{L_1}\quad(l=2,4,6,\dots),
\label{q2}
\eeq
with `$l$' for left.
The corresponding wavefunctions are standing waves living on the left ring,
whose amplitude vanishes at the contact point:\footnote{The symbol $\sim$ is
used whenever the wavefunction normalization is not given explicitly.}
\beq
\psi^\un(s_1)\sim\sin qs_1,\quad\psi^\de(s_2)=0.
\eeq

\item
{\it Right states}.
These states correspond to $\sin qL_2/2=0$, hence
\beq
q=\frac{r\pi}{L_2}\quad(r=2,4,6,\dots),
\label{q3}
\eeq
with `$r$' for right.
The corresponding wavefunctions are standing waves living on the right ring,
whose amplitude vanishes at the contact point:
\beq
\psi^\un(s_1)=0,\quad\psi^\de(s_2)\sim\sin qs_2.
\eeq

\end{itemize}

The ground-state energy, obtained by setting $b=0$ in~(\ref{q1}),
vanishes.\footnote{Notice that $l=0$ and $r=0$ are not allowed
in~(\ref{q2}) and~(\ref{q3}).}
The modulation introduced in~(\ref{gdef}) reads $g_1=-\pi$,
which saturates the bound~(\ref{gbound}).
The corresponding ground-state wavefunction is uniform:
$\psi^\un(s_1)=\psi^\de(s_2)=\psi(\PT)$.

For small fluxes, we have
\beq
E_1\approx\frac{1}{L_1+L_2}
\left(\frac{\Phi_1^2}{L_1}+\frac{\Phi_2^2}{L_2}\right),
\label{gse}
\eeq
so that the ground state is always diamagnetic.
As far as excited states are concerned,
the results~(\ref{q1}), (\ref{q2}) and~(\ref{q3}) respectively become
\beq
\matrix{
q(L_1+L_2)\approx b\pi+\frad{\Phi_1^2-\Phi_2^2}{2}\,\cot b\pi\o,\hfill\cr
qL_1\approx l\pi+\frad{\Phi_1^2}{2}\,\cot\frad{l\pi}{\o},\hfill\cr
qL_2\approx r\pi+\frad{\Phi_2^2}{2}\,\cot\frad{r\pi}{1-\o}.\hfill}
\label{qquadra}
\eeq
These results show that there is no general rule to predict
whether a given (left or right) state is paramagnetic or diamagnetic.
Bilateral states are always hybrid
(one paramagnetic ring and one diamagnetic one).

The only degeneracies between the three interlaced
spectra~(\ref{q1}),~(\ref{q2}) and~(\ref{q3})
are the threefold ones taking place in the commensurate case
at $qa=\theta=\pi\nu$, where $\nu=2\mu=2,4,6,\dots$ is an even integer.
With the notations of Section~\ref{com},
these momentum values correspond
to $b=\mu m$, $l=\mu m_1$ and~$r=\mu m_2$.
The quadratic terms in~(\ref{qquadra}) diverge.
This is in agreement with the fact that near a degeneracy
the energy eigenvalues vary linearly with the fluxes,
rather than quadratically (see Section~\ref{degene} for more details).

\subsection{Integer or half-integer fluxes}
\label{half}

Whenever each flux is either integer or half-integer
(i.e., an integer or a half-integer multiple of the flux quantum
$\Phi_0=2\pi$),
the spectrum of the system still consists of the above three types of states
(bilateral, left and right).
The corresponding momenta are still given
by~(\ref{q1}), (\ref{q2}) and~(\ref{q3}).
Table~\ref{tabinth} lists the characteristics of the spectrum
in the four different cases, corresponding to $\Phi_1=0$ or $\pi\mod{2\pi}$
and $\Phi_2=0$ or~$\pi\mod{2\pi}$.

\begin{table}[!ht]
\begin{center}
\begin{tabular}{|c|c||c|c|c|c|}
\hline
$\Phi_1$&$\Phi_2$&$D(q)$&$b$&$l$&$r$\\
\hline
\haut$0$&$0$&
$-4\,\sin\frad{q(L_1+L_2)}{2}\,\sin\frad{qL_1}{2}\,\sin\frad{qL_2}{2}$&
$0,2,4,\dots$&$2,4,6,\dots$&$2,4,6,\dots$\\
\hline
\haut$0$&$\pi$&
{\hskip 8pt}
$4\,\cos\frad{q(L_1+L_2)}{2}\,\sin\frad{qL_1}{2}\,\cos\frad{qL_2}{2}$&
$1,3,5,\dots$&$2,4,6,\dots$&$1,3,5,\dots$\\
\hline
\haut$\pi$&$0$&
{\hskip 8pt}
$4\,\cos\frad{q(L_1+L_2)}{2}\,\cos\frad{qL_1}{2}\,\sin\frad{qL_2}{2}$&
$1,3,5,\dots$&$1,3,5,\dots$&$2,4,6,\dots$\\
\hline
\haut$\pi$&$\pi$&
{\hskip 8pt}
$4\,\sin\frad{q(L_1+L_2)}{2}\,\cos\frad{qL_1}{2}\,\cos\frad{qL_2}{2}$&
$2,4,6,\dots$&$1,3,5,\dots$&$1,3,5,\dots$\\
\hline
\end{tabular}
\end{center}
\caption{Characteristics of the system for the four cases
with integer or half-integer fluxes:
factorized form of the characteristic function $D(q)$;
integers $b$, $l$ and $r$ entering the spectra~(\ref{q1}),
(\ref{q2}) and~(\ref{q3}),
respectively corresponding to bilateral, left and right states.}
\label{tabinth}
\end{table}

Left and right states are met in a more general setting.
More precisely, left states with even (resp.~odd) $l$ exist
as soon as $\Phi_1=0$ (resp.~$\Phi_1=\pi\mod{2\pi})$, irrespective of~$\Phi_2$,
whereas right states with even (resp.~odd) $r$ exist
as soon as $\Phi_2=0$ (resp.~$\Phi_2=\pi\mod{2\pi})$,
irrespective of $\Phi_1$.

\subsection{Quarter-integer fluxes}
\label{quarter}

The situation where each ring is threaded by a quarter flux unit
($\Phi_1=\Phi_2=\pi/2$) is also a special case of interest.
In this case, the characteristic function boils down to $D(q)=\sin q(L_1+L_2)$.
The momenta therefore have the linear dependence
\beq
q_n=\frac{n\pi}{L_1+L_2}\quad(n=1,2,\dots).
\label{linderiv}
\eeq

In spite of the simplicity of the spectrum,
the current amplitudes have the following non-trivial expressions:
\beq
Q_{1,n}=-\o\sin n\pi\o,\quad Q_{2,n}=(-1)^n(1-\o)\sin n\pi\o.
\label{qquarter}
\eeq
These quantities obey $\abs{Q_{1,n}}\le\o$, $\abs{Q_{2,n}}\le1-\o$,
and $\abs{Q_{1,n}}+\abs{Q_{2,n}}\le1$,
the latter bound being more stringent than the general one~(\ref{qbound}).
The currents in both rings therefore have the same sign (resp.~opposite signs)
whenever the level number $n$ is odd (resp.~even).

Using~(\ref{qalt}), the result~(\ref{qquarter}) can be recast as follows.
If the fluxes are close to a quarter flux quantum,
setting $\Phi_1=\pi/2+\delta\Phi_1$, $\Phi_2=\pi/2+\delta\Phi_2$,
the modulation $g_n$ introduced in~(\ref{gdef}) reads,
to first order in $\delta\Phi_1$ and $\delta\Phi_2$,
\beq
g_n\approx\sin n\pi\o\times
\left\{\matrix{
\delta\Phi_1-\delta\Phi_2&\hbox{for}&n\;\hbox{even},\cr
\delta\Phi_1+\delta\Phi_2&\hbox{for}&n\;\hbox{odd}.\hfill
}\right.%}
\label{gfirst}
\eeq

\section{Degeneracies}
\label{degene}

A degeneracy manifests itself as a multiple (i.e., at least double) root
of the characteristic equation~(\ref{dzero}),
i.e., as a simultaneous root of $D(q)=0$ and $\dpar D/\dpar q=0$.
The bound~(\ref{derbound}) shows that
{\it the system has no accidental degeneracy.}
A degeneracy may indeed only take place when both products
$\sin\Phi_1\,\sin qL_2$ and $\sin\Phi_2\,\sin qL_1$
vanish simultaneously, i.e., when at least one factor of each product vanishes.
It can be checked that there are only two types of degeneracies,
to be successively investigated hereafter.

\subsection{Twofold degeneracies}
\label{degene2}

Twofold degeneracies occur when either
$\sin qL_1=\sin\Phi_1=0$ (but $\sin qL_2\ne0$)
or $\sin qL_2=\sin\Phi_2=0$ (but $\sin qL_1\ne0$).
These two instances will be respectively referred to
as left and right degeneracies, for a reason that will become clear in a while.

Consider the first instance for definiteness.
This is a left degeneracy,
as the state to become twofold degenerate is a left state,
in the sense of Section~\ref{absence}.
We have $\cos qL_1=\cos\Phi_1=(-1)^l$,
with the notation~(\ref{q2}), so that
\beq
\frac{\dpar D}{\dpar q}=(-1)^l L_1(\cos qL_2-\cos\Phi_2).
\eeq
The conditions for a twofold degeneracy are therefore
\beq
\sin qL_1=\sin\Phi_1=0,\quad\cos qL_2=\cos\Phi_2.
\label{dege2}
\eeq
At the degeneracy the momentum is $q=l\pi/L_1$,
whereas the fluxes read $\Phi_1=(2j+l)\pi$ and
$\Phi_2=(2k+\eps lL_2/L_1)\pi$, where $\eps=\pm1$,
whereas $j$ and $k$ are integers.
For circular rings (more generally, for rings with similar shapes),
we have $\Phi_2/\Phi_1=A_2/A_1=(L_2/L_1)^2=((1-\o)/\o)^2$.
This condition relates $\o$ to $\eps$, $j$ and $k$ as follows:
\beq
\matrix{
\eps=+1:\hfill&2(j-k+l)\o^2-(4j+3l)\o+2j+l=0,\hfill\cr
\eps=-1:\hfill&2(j-k)\o^2-(4j+l)\o+2j+l=0.\hfill}
\eeq
The system may therefore have twofold degeneracies
both in the commensurate case ($\o$ rational)
and in the incommensurate case provided $\o$ is a quadratic number,
obeying an equation of the form
\beq
I\o^2+J\o+K=0,
\label{wquadra}
\eeq
where $I$, $J$ and $K$ are integers.

In order to explore the vicinity of the degeneracy, we set
\beq
q=\frac{l\pi}{L_1}+\eta,\quad
\Phi_1=l\pi+\delta\Phi_1,\quad
\Phi_2=\frac{l\pi L_2}{L_1}+\delta\Phi_2.
\eeq
By expanding the characteristic equation~(\ref{dzero})
to second order in $\eta$,
$\delta\Phi_1$ and $\delta\Phi_2$, we obtain the reduced equation
\beq
L_1(L_1+2L_2)\eta^2-2L_1\delta\Phi_2\,\eta-\delta\Phi_1^2=0.
\eeq
The momenta of the two nearly degenerate states are therefore given by
\beq
\eta_\pm=\frac{L_1\delta\Phi_2\pm W_2}{L_1(L_1+2L_2)},
\eeq
with the definition
\beq
W_2=\left(L_1(L_1+2L_2)\delta\Phi_1^2+L_1^2\delta\Phi_2^2\right)^{1/2}.
\eeq
The momenta shifts $\eta_\pm$ vanish linearly with the distance
to the degeneracy in the $(\delta\Phi_1,\delta\Phi_2)$ plane.
The corresponding amplitudes of the persistent currents read
\beq
Q_{1,\pm}=\mp\frac{L_1\delta\Phi_1}{W_2},\quad
Q_{2,\pm}=\frac{L_2}{L_1+2L_2}\left(1\pm\frac{L_1\delta\Phi_2}{W_2}\right).
\eeq
These amplitudes therefore remain of order unity,
and they depend on the direction in the $(\delta\Phi_1,\delta\Phi_2)$ plane
along which the degeneracy point is approached.
As this direction is varied, the amplitudes describe an ellipse:
\beq
(1-\o)^2Q_1^2+\o(2-\o)Q_2^2-2\o(1-\o)Q_2=0.
\label{e2}
\eeq

\subsection{Threefold degeneracies}
\label{degene3}

We have noticed at the end of Section~\ref{absence} that the spectrum has
threefold degeneracies in the absence of fluxes
in the commensurate case at $qa=\theta=\pi\nu$,
where $\nu=2\mu=2,4,\dots$ is an even integer,
with the notations of Section~\ref{com}.

More generally, there are threefold degeneracies at all the energies
of the idle states, given by~(\ref{idle}),
i.e., $qa=\theta=\pi\nu$, with $\nu=1,2,\dots$
Consider indeed one of these energies.
We have
\beq
\frac{\dpar D}{\dpar q}=a\eps_1\eps_2
(m_1(1-\eps_2\,\cos\Phi_2)+m_2(1-\eps_1\,\cos\Phi_1)),
\eeq
with the notations~(\ref{eps12}).
The conditions for a threefold degeneracy are therefore
\beq
\cos\Phi_1=\eps_1,\quad\cos\Phi_2=\eps_2.
\label{cond3}
\eeq
Right at the degeneracy, the fluxes therefore read
$\Phi_1=\pi\nu m_1$ and $\Phi_2=\pi\nu m_2\mod{2\pi}$.
The situation in the absence of fluxes is recovered when $\nu$ is even.
If $\nu$ is odd, at least one of the fluxes is a half integer,
because at least one of the integers $m_1$, $m_2$ is odd.

In order to explore the vicinity of the degeneracy, we set
\beq
\theta=\nu\pi+\eta,\quad
\Phi_1=\pi\nu m_1+\delta\Phi_1,\quad
\Phi_2=\pi\nu m_2+\delta\Phi_2.
\eeq
By expanding the characteristic equation~(\ref{dzero}) to third order in $\eta$,
and to second order in $\delta\Phi_1$ and $\delta\Phi_2$,
we find three solutions:
the idle state at $\eta=0$,
and two symmetrically shifted active states at $\eta_\pm=\pm W_3$, with
\beq
W_3=
\left(\frac{m_2\delta\Phi_1^2+m_1\delta\Phi_2^2}{m_1m_2(m_1+m_2)}\right)^{1/2}.
\eeq
The momentum shift $W_3$ again vanishes linearly with the distance
to the degeneracy in the ($\delta\Phi_1,\delta\Phi_2$) plane.
The corresponding amplitudes of the persistent currents read
\beq
Q_{1,\pm}=\mp\frac{\delta\Phi_1}{(m_1+m_2)W_3},\quad
Q_{2,\pm}=\mp\frac{\delta\Phi_2}{(m_1+m_2)W_3}.
\eeq
Here again, these amplitudes remain of order unity
and depend on the direction in the ($\delta\Phi_1,\delta\Phi_2$) plane
along which the degeneracy point is approached.
As this direction is varied, the amplitudes describe an ellipse:
\beq
(1-\o)Q_1^2+\o Q_2^2-\o(1-\o)=0.
\label{e3}
\eeq

\section{Commensurate ring lengths}
\label{com}

We now turn to the situation where the ring lengths $L_1$ and $L_2$
are commensurate.
One of the peculiar features of this situation is the existence
of threefold degeneracies, already investigated in Section~\ref{degene3}.

In the commensurate situation, both ring lengths are multiples
of the same fundamental length $a$:
\beq
L_1=m_1a,\quad L_2=m_2a,
\eeq
where the integers $m_1\ge1$ and $m_2\ge1$ are relatively prime.
Setting $m=m_1+m_2\ge2$, the variable $\o$ takes the rational value $\o=m_1/m$.
For circular commensurate rings,
the fluxes read $\Phi_1=m_1^2\,Ba^2/(4\pi)$, $\Phi_2=m_2^2\,Ba^2/(4\pi)$,
so that the magnetization is periodic in the magnetic field $B$,
with period $B_0=8\pi^2/a^2$.

\subsection{Spectrum}

Introducing the reduced momentum
\beq
\theta=qa,
\eeq
the characteristic function becomes
\beq
D(\theta)=\sin m\theta-\cos\Phi_2\,\sin m_1\theta-\cos\Phi_1\,\sin m_2\theta.
\eeq
The spectrum of the system is $2\pi$-periodic in $\theta$.
It consists of two types of states.

\begin{itemize}

\item
{\it Idle states.}
They correspond to the trivial solutions of the characteristic equation:
\beq
\theta=\pi\nu\quad(\nu=1,2,\dots).
\label{idle}
\eeq
These states carry no current, as their energy is independent of the fluxes.
The corresponding wavefunctions do however depend on the fluxes.
We have indeed
\beq
\matrix{
\psi^\un(s_1)\sim
(\e^{-\i\Phi_2/2}-\eps_2\,\e^{\i\Phi_2/2})\e^{\i a_1s_1}\sin qs_1,\cr
\psi^\de(s_2)\sim
(\eps_1\,\e^{\i\Phi_1/2}-\e^{-\i\Phi_1/2})\e^{\i a_2s_2}\sin qs_2,}
\eeq
with
\beq
\eps_1=\e^{\i qL_1}=(-1)^{\nu m_1},\quad\eps_2=\e^{\i qL_2}=(-1)^{\nu m_2}.
\label{eps12}
\eeq
Idle states become threefold degenerate for integer or half-integer
values of the fluxes such that the condition~(\ref{cond3})
is fulfilled.

\item
{\it Active states.}
These states, corresponding to the other solutions
of the char\-ac\-te\-r\-istic equation $D(\theta)=0$,
carry non-zero persistent currents in general.

Setting $c=\cos\theta$, we have the identity
$\sin m\theta=\sin\theta\,U_{m-1}(c)$,
where $U_n(c)$ is the $n$th Tchebyshev polynomial of the second kind,
whose degree is $n$.
Focusing onto active states, the characteristic equation therefore reads
\beq
U_{m-1}(c)-\cos\Phi_2\,U_{m_1-1}(c)-\cos\Phi_1\,U_{m_2-1}(c)=0.
\label{uchar}
\eeq
This equation has $m-1$ solutions $c_{(k)}$ ($k=1,\dots,m-1$).
We write these solutions as $c_{(k)}=\cos\theta_{(k)}$,
with $0\le\theta_{(k)}\le\pi$, ordered as
\beq
0\le\theta_{(1)}\le\dots\le\theta_{(m-1)}\le\pi.
\eeq
Threefold degeneracies correspond to the limiting situations
$\theta_{(1)}=0$ or $\theta_{(m-1)}=\pi$,
whereas twofold ones can take place anywhere along the sequence
of $\theta_{(k)}$'s.

\end{itemize}

To sum up, in each period of the spectrum of length $2\pi$
in the variable $\theta=qa$,
there are $2(m-1)$ active levels and~2 idle ones, i.e., a total of $2m$ states.
The modulation and the current amplitudes
associated with the states at $\theta=\theta_{(k)}$ ($k=1,\dots,m-1$) read
\beq
g_k=m\theta_{(k)}-k\pi,\quad
Q_{1,k}=-m_1\,\frac{\dpar\theta_{(k)}}{\dpar\Phi_1},\quad
Q_{2,k}=-m_2\,\frac{\dpar\theta_{(k)}}{\dpar\Phi_2},
\label{qc1}
\eeq
whereas those associated with the states at $\theta=2\pi-\theta_{(i)}$,
\beq
g_{2m-k}=k\pi-m\theta_{(k)},\quad
Q_{1,2m-k}=m_1\,\frac{\dpar\theta_{(k)}}{\dpar\Phi_1},\quad
Q_{2,2m-k}=m_2\,\frac{\dpar\theta_{(k)}}{\dpar\Phi_2},
\label{qc2}
\eeq
are the opposites of the first expressions.
The bound~(\ref{gbound}) implies
\beq
(k-1)\pi\le m\theta_{(k)}\le(k+1)\pi\quad(k=1,\dots,m-1).
\eeq
Modulation and current amplitudes then repeat themselves periodically,
with period~$2m$.
The main characteristics of the states in the first period
are listed in Table~\ref{tablevels}.

\begin{table}[!ht]
\begin{center}
\begin{tabular}{|c||c|c|c|c|c|c|c|c|}
\hline
$n$&1&$\dots$&$m-1$&$m$&$m+1$&$\dots$&$2m-1$&$2m$\\
\hline
$\theta$&$\theta_{(1)}$&$\dots$&$\theta_{(m-1)}$&$\pi$&
$2\pi-\theta_{(m-1)}$&$\dots$&$2\pi-\theta_{(1)}$&$2\pi$\\
\hline
type&A&$\dots$&A&I&A&$\dots$&A&I\\
\hline
$g_n$&$g_1$&$\dots$&$g_{m-1}$&0&$-g_{m-1}$&$\dots$&$-g_1$&0\\
\hline
$Q_{1,n}$&$Q_{1,1}$&$\dots$&$Q_{1,m-1}$&0&$-Q_{1,m-1}$&$\dots$&$-Q_{1,1}$&0\\
\hline
$Q_{2,n}$&$Q_{2,1}$&$\dots$&$Q_{2,m-1}$&0&$-Q_{2,m-1}$&$\dots$&$-Q_{2,1}$&0\\
\hline
\end{tabular}
\end{center}
\caption{The first $2m$ states of the system in the commensurate case,
corresponding to the first period of the spectrum in the variable $\theta=qa$.
For each column, corresponding to a state,
the table gives the level number $n$, the angle $\theta$,
the type of state (`I' for idle or `A' for active),
the modulation and the current amplitudes.}
\label{tablevels}
\end{table}

\subsection{Rings with equal lengths}

The case where both rings have equal lengths
is the simplest of all the commensurate situations.
We have $m_1=m_2=1$, $m=2$, $\o=1/2$ and $a=L=L_1=L_2$.
The characteristic equation~(\ref{uchar}) has a single solution:
\beq
c_{(1)}=\cos\theta_{(1)}=\frac{\cos\Phi_1+\cos\Phi_2}{2}.
\eeq
Idle states ($n=2p$) and active states ($n=2p-1$) alternate along the spectrum:
\beq
q_{2p}=\frac{p\pi}{L},\quad q_{2p-1}=\frac{(2p-1)\pi+(-1)^{p-1}g_1}{2L}.
\label{altern}
\eeq
The modulation and the current amplitudes of the lowest active state
($n=2p-1=1$, i.e., $p=1$) read
\beq
g_1=2\theta_{(1)}-\pi,\quad
Q_{1,1}=-\frac{\sin\Phi_1}{2\sin\theta_{(1)}},\quad
Q_{2,1}=-\frac{\sin\Phi_2}{2\sin\theta_{(1)}}.
\eeq

In the case of two equal circular rings,
the above predictions can be made fully explicit.
We have $\Phi_1=\Phi_2=\Phi=BL^2/(4\pi)$.
Symmetry considerations allow us to restrict ourselves to $0\le\Phi\le\pi$,
so that $\theta_{(1)}=\Phi$, $g_1=2\Phi-\pi$ and $Q_{1,1}=Q_{2,1}=-1/2$.
Hence, using~(\ref{im}),
the contribution of any odd state ($n=2p-1$) to the magnetization reads
\beq
M_{2p-1}=(-1)^p\,\frac{2p-1}{4}+\frac{\pi-2\Phi}{4\pi}.
\eeq
Inserting this formula into~(\ref{total}),
we obtain the expression of the magnetization
of a system with $N$ electrons at zero temperature.
This expression, given in Table~\ref{tabtotal}, depends on $N\mod{4}$.
The corresponding result for a single circular ring,
given by~(\ref{sodd}),~(\ref{seven}), is also listed in the Table for
comparison
(the latter depends on the sign of $N$, i.e., on~$N\mod{2}$).

\begin{table}[!ht]
\begin{center}
\begin{tabular}{|c||r|r|}
\hline
$N\mod{4}$&$M$ (two equal rings)&$M$ (one single ring)\\
\hline
0&$N(\pi-\Phi)/(4\pi)$&$N(\pi-\Phi)/(2\pi)$\\
\hline
1&$-(N+1)\Phi/(4\pi)$&$-N\Phi/(2\pi)$\\
\hline
2&$-N\Phi/(4\pi)$&$N(\pi-\Phi)/(2\pi)$\\
\hline
3&$(N+1)(\pi-\Phi)/(4\pi)$&$-N\Phi/(2\pi)$\\
\hline
\end{tabular}
\end{center}
\caption{Expression of the magnetization $M$
of a system with $N$ electrons at zero temperature,
as a function of the flux $\Phi$ in the range $0\le\Phi\le\pi$,
for two equal circular rings and for a single circular ring.}
\label{tabtotal}
\end{table}

\subsection{Magnetization}

The simple linear dependence of the magnetization on the magnetic field
in the range $0\le\Phi\le\pi$, observed in Table~\ref{tabtotal},
is a peculiarity of the cases considered there,
namely one single circular ring or two equal ones.

\begin{figure}[!ht]
\begin{center}
\includegraphics[angle=90,width=.4\linewidth]{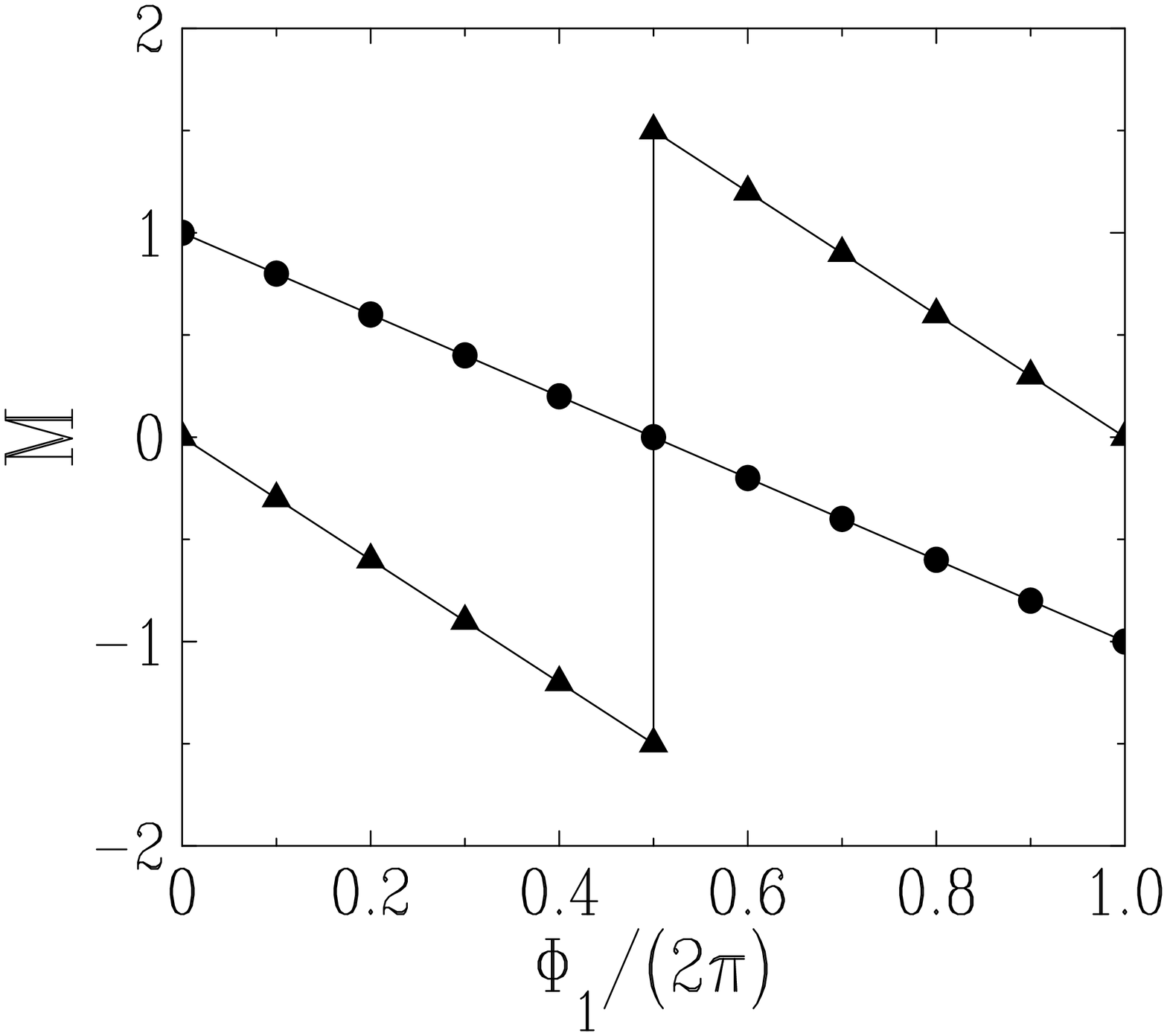}
{\hskip 10pt}
\includegraphics[angle=90,width=.4\linewidth]{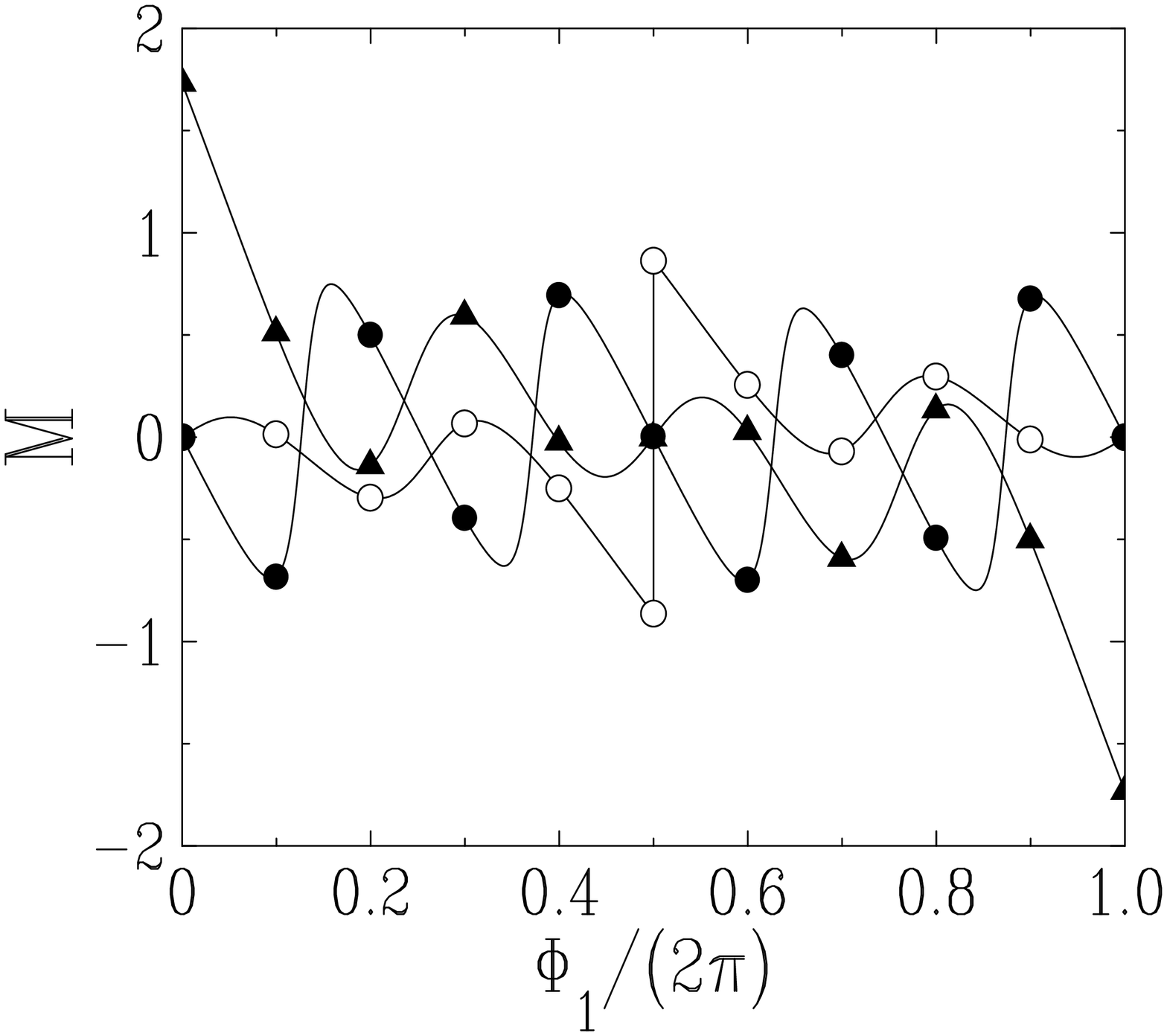}
\caption{Plots of the magnetization $M$ over one period,
against $\Phi_1/(2\pi)=BL_1^2/(8\pi^2)$,
for commensurate circular rings
with $L_2=L_1$ (left) and $L_2=2L_1$ (right),
for $N=3$ (empty circles), $N=4$ (full circles) and $N=5$ (full triangles).
Data for $N=3$ and $N=4$ coincide on the left plot, as $n=4$ is an idle state.}
\label{figmb}
\end{center}
\end{figure}

In all the other cases of commensurate circular rings,
the curve $M(B)$ is non-trivial.
This is illustrated in Figure~\ref{figmb},
showing plots of the magnetization over one period,
against $\Phi_1/(2\pi)=B/B_0=BL_1^2/(8\pi^2)$,
for $L_2=L_1$ (left) and $L_2=2L_1$ (right).
The latter example ($m_1=1$, $m_2=2$, $m=3$)
is generic of the commensurate case,
except that the characteristic equation~(\ref{uchar})
is of degree two in $c$ and can therefore be solved analytically.
We thus obtain
\beq
\matrix{
c_{(1)}=\cos\theta_{(1)}
=\frad{1}{4}\left(\cos\Phi_1+(4+4\cos\Phi_2+\cos^2\Phi_1)^{1/2}\right),\cr
c_{(2)}=\cos\theta_{(2)}
=\frad{1}{4}\left(\cos\Phi_1-(4+4\cos\Phi_2+\cos^2\Phi_1)^{1/2}\right).}
\eeq
The currents and the magnetization then follow from
(\ref{qc1}), (\ref{qc2}) and (\ref{im}), (\ref{total}).

\section{Incommensurate ring lengths}
\label{incom}

We now turn to the situation where the ring lengths $L_1$ and $L_2$
are incommensurate, i.e., the variable $\o$ is irrational.

The key quantity is again the modulation $g_n$ of the spectrum.
In the commensurate case, i.e., for a rational $\o=m_1/m$,
$g_n$ has been shown to be periodic in $n$, with period~$2m$.
In the present incommensurate case,
$g_n$ is expected to never repeat itself exactly.
At a quantitative level, the structure of $g_n$ is revealed
by the situation where $\Phi_1=\Phi_2=\pi/2$,
considered in Section~\ref{quarter}.
The result~(\ref{gfirst}) suggests the Ansatz
\beq
\matrix{
n\;\hbox{even}:\hfill&g_n=\gp(n\pi\o),\hfill\cr
n\;\hbox{odd}:\hfill&g_n=\gi(n\pi\o),\hfill}
\label{Ansatz}
\eeq
where $\gp$ (`e' for even) and $\gi$ (`o' for odd)
are $2\pi$-periodic functions of $x=n\pi\o$.
The modulation is therefore
a {\it quasiperiodic} function of the level number $n$.
We will refer to~$\gp$ and $\gi$ as the {\it hull functions},
following the term introduced by Aubry in the context
of modulated incommensurate structures~\cite{hull}.
The current amplitudes are then also quasiperiodic functions of $n$:
\beq
\matrix{
n\;\hbox{even}:\hfill&
Q_{1,n}=-\o\,\frad{\dpar\gp(n\pi\o)}{\dpar\Phi_1},\hfill&
Q_{2,n}=-(1-\o)\,\frad{\dpar\gp(n\pi\o)}{\dpar\Phi_2},\cr
n\;\hbox{odd}:\hfill&
Q_{1,n}=-\o\,\frad{\dpar\gi(n\pi\o)}{\dpar\Phi_1},\hfill&
Q_{2,n}=-(1-\o)\,\frad{\dpar\gi(n\pi\o)}{\dpar\Phi_2}.}
\label{qqp}
\eeq

Inserting~(\ref{Ansatz}) into the characteristic equation~(\ref{dzero})
leads to implicit equations for the hull functions:
\beq
\matrix{
\sin\gp=\cos\Phi_2\sin(x+\o\gp)-\cos\Phi_1\sin(x-(1-\o)\gp),\hfill\cr
\sin\gi=-\cos\Phi_2\sin(x+\o\gi)-\cos\Phi_1\sin(x-(1-\o)\gi).}
\label{hulleqs}
\eeq
These equations imply the following properties:
$\gp(x)$ and $\gi(x)$ are $2\pi$-periodic, odd and continuous functions of $x$;
$\gp(x)=0$ for $\cos\Phi_1=\cos\Phi_2$,
whereas $\gi(x)=0$ for $\cos\Phi_1=-\cos\Phi_2$;
changing $\cos\Phi_2$ into its opposite amounts to exchanging $\gp$ and~$\gi$,
whereas changing $\cos\Phi_1$ and $\cos\Phi_2$ into their opposites
amounts to changing $x$ into~$x+\pi$.

\subsection{Integer or half-integer fluxes}

It is worth investigating first the case where there are no magnetic fluxes,
and more generally the situation where the fluxes are integer or half-integer.
The spectrum in the absence of magnetic fluxes
has been studied in Section~\ref{absence}.

\begin{itemize}

\item
Even values $n=2p$ of the level number correspond to bilateral states.
The expression~(\ref{q1}) shows that the modulation vanishes
whenever $n=b=2p$ is an even integer.
We have therefore $\gp=0$.

\item
Odd values $n=2p-1$ of the level number correspond to left and right states.
For left states, setting $l=2\lbar$, ~(\ref{q2}) yields
$2(p-1)\pi<(2p-1)\pi+g_{2p-1}=2\lbar\pi/\o<2p\pi$,
hence\footnote{We recall that the {\it integer part} $\Int x$
and the {\it fractional part} $\Frac x$ of a real number $x$
are defined by $x=\Int x+\Frac x$,
with $\Int x$ integer and $0\le\Frac x<1$,
so that $\Frac x$ is periodic in $x$, with unit period.}
\beq
p=1+\Int\frac{\lbar}{\o},\quad
g_{2p-1}=2\pi\left(\Frac\frac{\lbar}{\o}-\frac{1}{2}\right).
\label{pgl}
\eeq
Similarly, for right states, setting $r=2\rbar$,~(\ref{q3}) yields
$2(p-1)\pi<(2p-1)\pi+g_{2p-1}=2\rbar\pi/(1-\o)<2p\pi$,
hence
\beq
p=1+\Int\frac{\rbar}{1-\o},\quad
g_{2p-1}=2\pi\left(\Frac\frac{\rbar}{1-\o}-\frac{1}{2}\right).
\label{pgr}
\eeq

\end{itemize}

The expressions~(\ref{pgl}) and~(\ref{pgr}) cover every integer $p$ once.
The inverse formulas read
\beq
\matrix{
\Frac p\o<\o:\hfill&\lbar=\Int p\o,\hfill&
g_{2p-1}=\pi-\frad{2\pi\,\Frac p\o}{\o},\hfill\cr
\Frac p\o>\o:\hfill&\rbar=p-1-\Int p\o,\hfill&
g_{2p-1}=\pi-\frad{2\pi(1-\Frac p\o)}{1-\o}.\hfill}
\eeq
The latter expressions for the modulation
can be brought to the form~(\ref{Ansatz}) with $\gi(x)=\G(x)$,
the periodic, odd, piecewise linear continuous function
defined for $0\le x\le 2\pi$ as:
\beq
\G(x)=\left\{\matrix{
-x/\o\hfill&\hbox{for}&0\le x\le\pi\o,\hfill\cr
(x-\pi)/(1-\o)\hfill&\hbox{for}&\pi\o\le x\le\pi(2-\o),\hfill\cr
(2\pi-x)/\o\hfill&\hbox{for}&(2-\o)\pi\le x\le2\pi.\hfill\cr
}\right.%}
\label{Gdef}
\eeq
The linearly increasing (resp.~decreasing) parts of $\G(x)$
describe right (resp.~left) states.
The cusps at $x=\pi\o$ and $x=\pi(2-\o)$
eventually correspond to threefold degeneracies.

\begin{figure}[!ht]
\begin{center}
\includegraphics[angle=90,width=.4\linewidth]{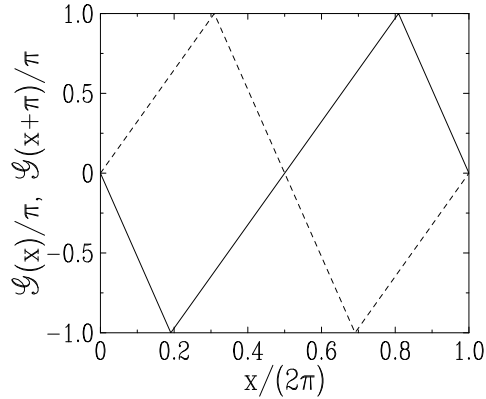}
\caption{Plot of the hull functions
$\G(x)$ (full line) and $\G(x+\pi)$ (dashed line), divided by $\pi$,
against $x/(2\pi)$ over one period, for $\o=1/\tau^2$.}
\label{figgg}
\end{center}
\end{figure}

More generally, for the four situations corresponding either
to integer or half-integer fluxes, considered in Section~\ref{half},
the non-zero hull functions are either equal to $\G(x)$ or to $\G(x+\pi)$,
as shown in Table~\ref{tabhulls}.
Figure~\ref{figgg} shows a plot of these two functions in
the prototypical example of an incommensurate situation,
namely $L_2/L_1=\tau$, where $\tau=(1+\sqrt5)/2\approx1.618\,034$
is the Golden mean, so that $\o=1/\tau^2=(3-\sqrt5)/2\approx0.381\,966$.

\begin{table}[!ht]
\begin{center}
\begin{tabular}{|c|c||c|c|}
\hline
$\Phi_1$&$\Phi_2$&$\gp(x)$&$\gi(x)$\\
\hline
$0$&$0$&$0$&$\G(x)$\\
\hline
$0$&$\pi$&$\G(x)$&$0$\\
\hline
$\pi$&$0$&$\G(x+\pi)$&$0$\\
\hline
$\pi$&$\pi$&$0$&$\G(x+\pi)$\\
\hline
\end{tabular}
\end{center}
\caption{Expression of the hull functions $\gp(x)$ and $\gi(x)$,
respectively characterizing the states with even and odd $n$,
for the four cases with integer or half-integer fluxes.
The function $\G(x)$ is defined in~(\ref{Gdef}).}
\label{tabhulls}
\end{table}

\subsection{The general case}

Coming back to the general case, i.e., arbitrary values of the fluxes,
we are now in a position to show another remarkable property:
the hull functions
are bounded by their limiting values for integer or half-integer fluxes:
\beq
\matrix{
0\le x\le\pi:\hfill&
\G(x)\le\gp(x),\;\gi(x)\le\G(x+\pi),\cr
\pi\le x\le2\pi:\hfill&
\G(x+\pi)\le\gp(x),\;\gi(x)\le\G(x).}
\label{hullineqs}
\eeq
In other words, the hull functions are inscribed in the two parallelograms
shown in Figure~\ref{figgg}.
This property implies in particular
$\abs{\gp(x)}\le\pi$ and $\abs{\gi(x)}\le\pi$, hence the bound~(\ref{gbound}).

\begin{figure}[!ht]
\begin{center}
\includegraphics[angle=90,width=.4\linewidth]{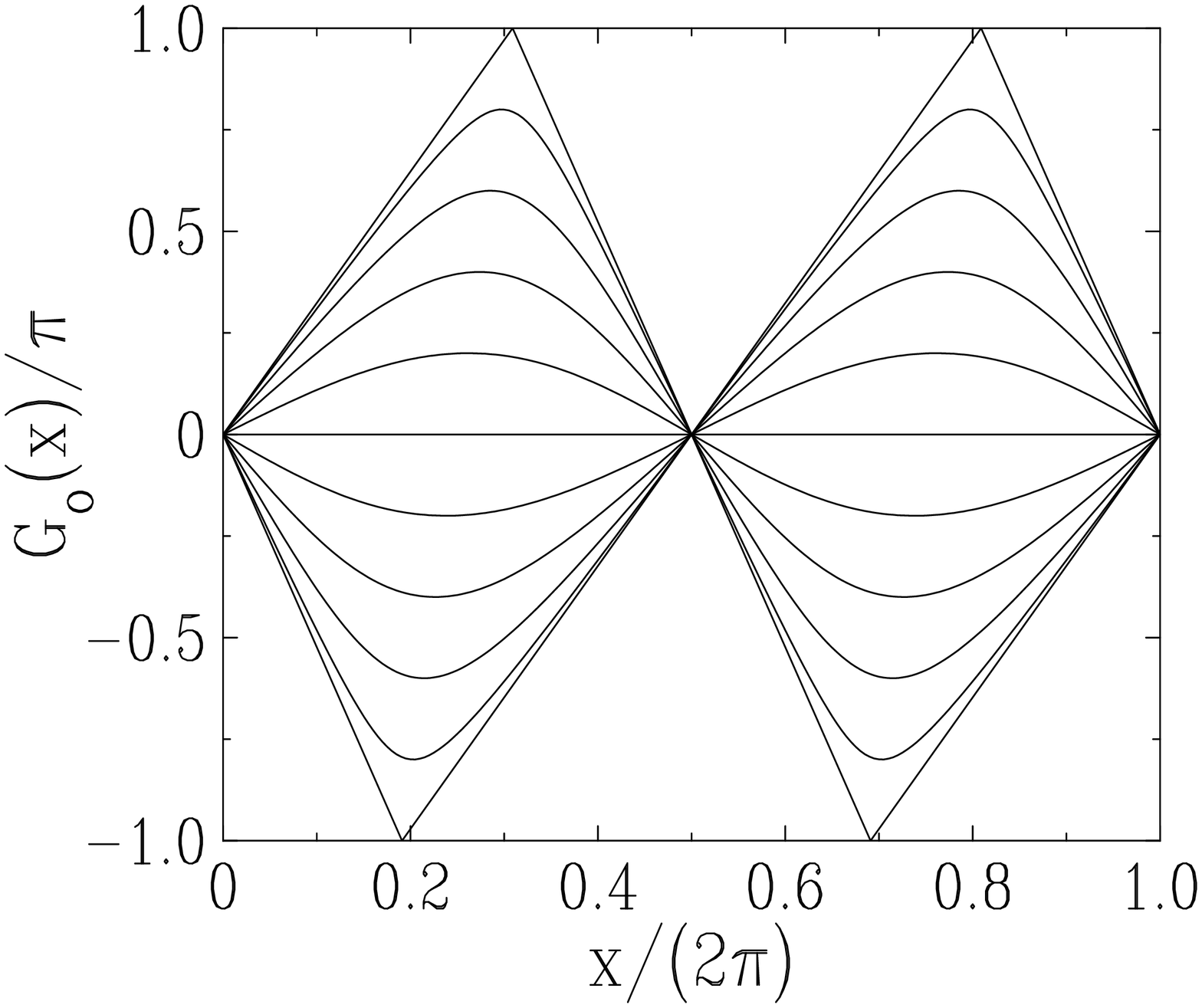}
{\hskip 10pt}
\includegraphics[angle=90,width=.4\linewidth]{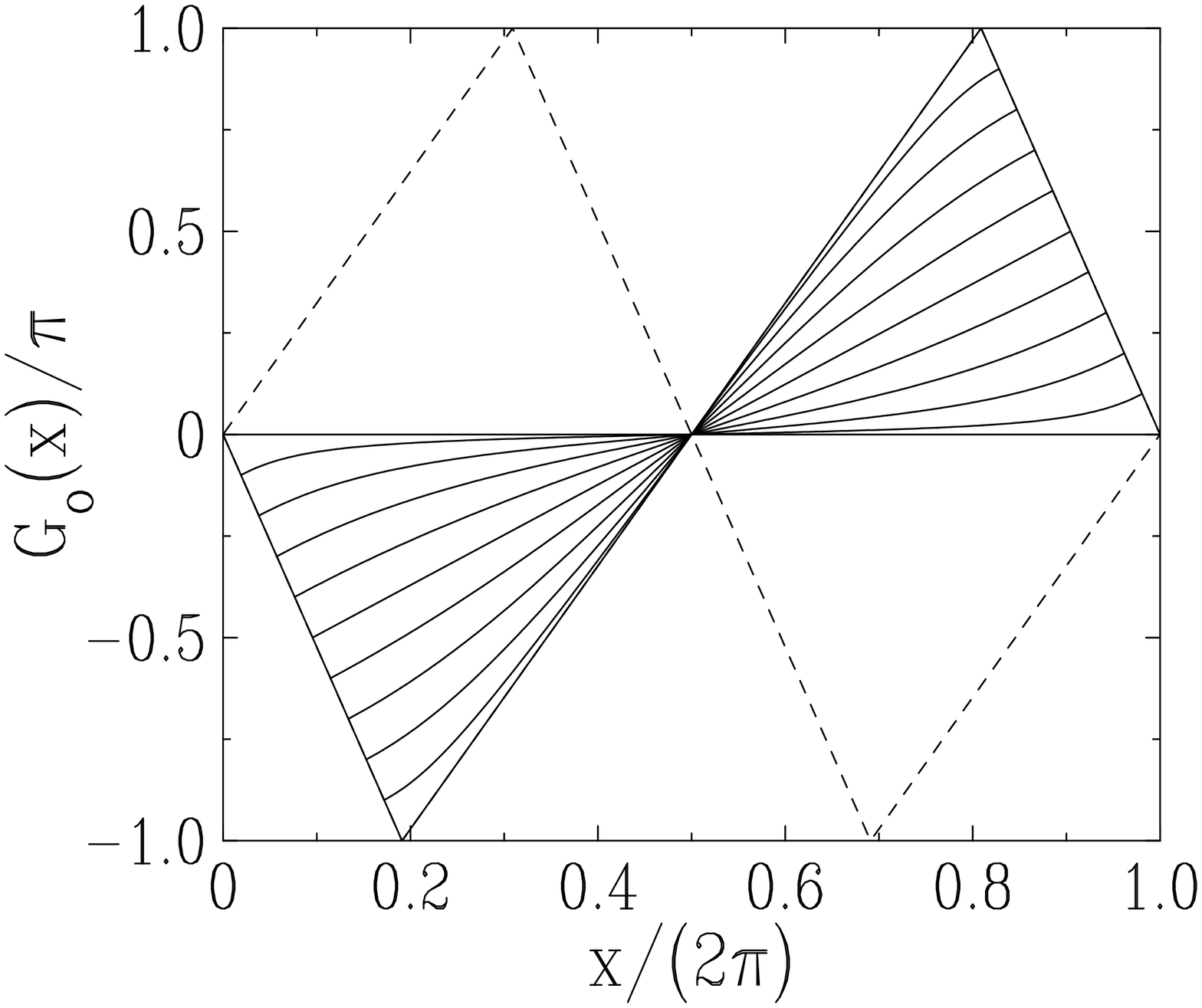}
\caption{Plots of the hull function $\gi(x)$, divided by $\pi$,
against $x/(2\pi)$ over one period, for $\o=1/\tau^2$.
Left panel: $\Phi_1=\Phi_2=k\pi/10$.
Right panel: $\Phi_1=0$, $\Phi_2=k\pi/10$.
The latter case exhibits the linear parts~(\ref{glin}).
In both cases, $k=0,\dots,10$, bottom to top in the left part of the curves.}
\label{figg}
\end{center}
\end{figure}

The inequalities~(\ref{hullineqs}) can be proved as follows.
It can be checked, using their expressions~(\ref{qqp}),
that the current amplitudes $Q_{1,n}$ and $Q_{2,n}$
have well-defined signs:
\beq
\matrix{
\sign(I_{1,n})=\sign(Q_{1,n})=-\sign(\sin\Phi_1)\sign(\sin n\pi\o),\hfill\cr
\sign(I_{2,n})=\sign(Q_{2,n})=-\sign(\sin\Phi_2)
\underbrace{\sign(\sin n\pi(1-\o))}_{(-1)^{n-1}\sign(\sin n\pi\o)},}
\eeq
as long as both $\sin\Phi_1$ and $\sin\Phi_2$ are non-zero,
so that degeneracies are avoided.
These signs are constant over the domain $0<\Phi_1<\pi$, $0<\Phi_2<\pi$.
In particular, the observation made in Section~\ref{quarter},
that the currents in both rings have the same sign (resp.~opposite signs)
whenever the level number $n$ is odd (resp.~even),
holds all over this domain.
The hull functions therefore have a monotonic dependence
on $\Phi_1$ and $\Phi_2$ in the same domain,
and they take their extremal values at the corners of the domain,
i.e., for integer or half-integer fluxes.
Figure~\ref{figg} shows plots of the hull function $\gi$ for $\o=1/\tau^2$,
both for $\Phi_1=\Phi_2$ (left)
and for $\Phi_1=0$ and variable $\Phi_2$ (right).

If one flux is integer or half-integer, albeit the other is not,
the hull functions $\gp(x)$ or $\gi(x)$ exhibit linear parts,
where they coincide either with $\G(x)$ or with $\G(x+\pi)$.
These linear parts describe left or right states.
They end at cusps which eventually correspond to twofold degeneracies.
Consider for definiteness $\Phi_1=0$ and $0<\Phi_2<\pi$.
The hull functions start linearly as:
\beq
\matrix{
n\;\hbox{even}:\hfill&\gp(x)=-x/\o&\hbox{for}&0\le x\le\xp=\o\Phi_2,\hfill\cr
n\;\hbox{odd}:\hfill&\gi(x)=-x/\o&\hbox{for}&0\le x\le\xi=\o(\pi-\Phi_2).}
\label{glin}
\eeq

\begin{figure}[!ht]
\begin{center}
\includegraphics[angle=90,width=.4\linewidth]{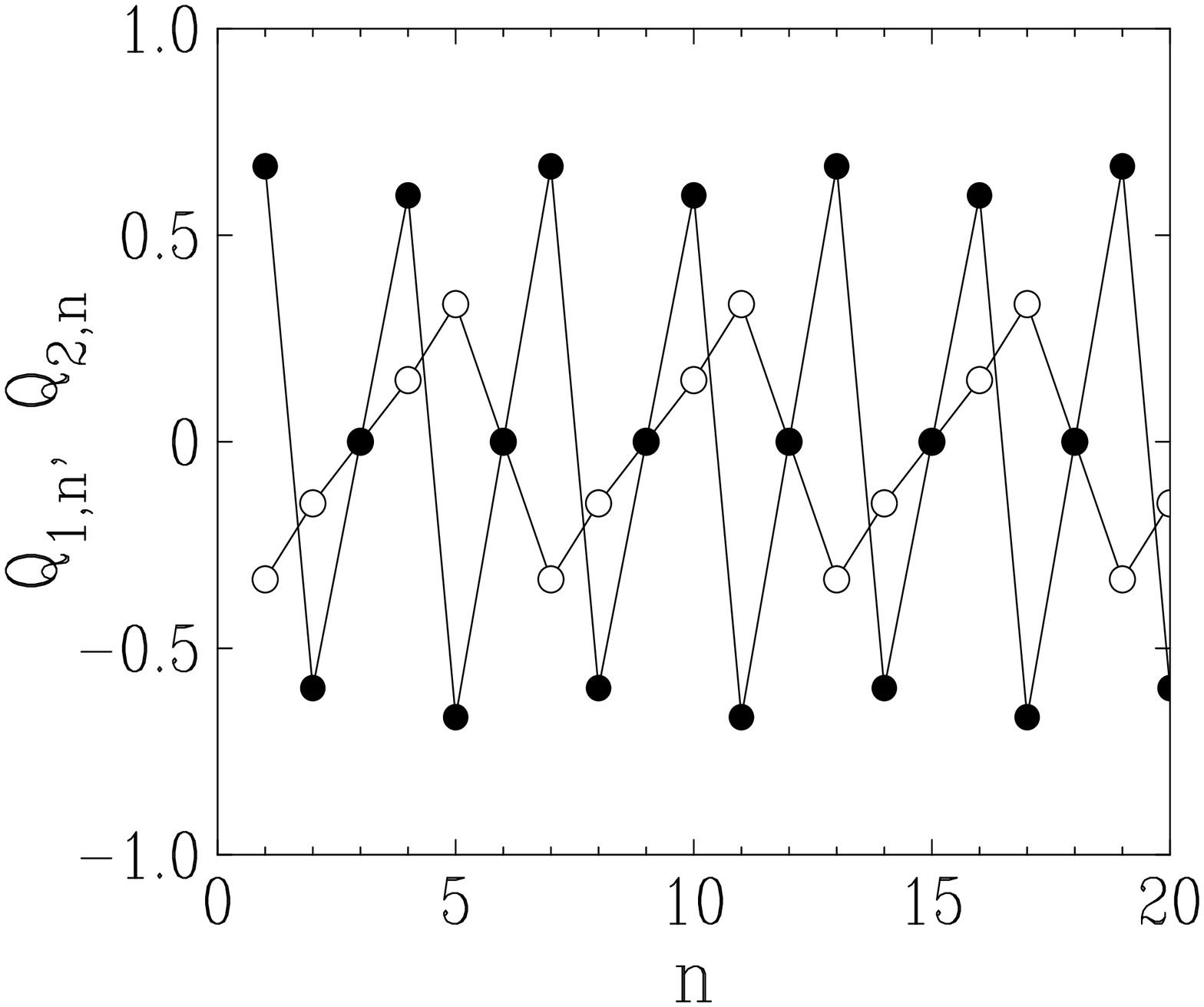}
{\hskip 0pt}
\includegraphics[angle=90,width=.4\linewidth]{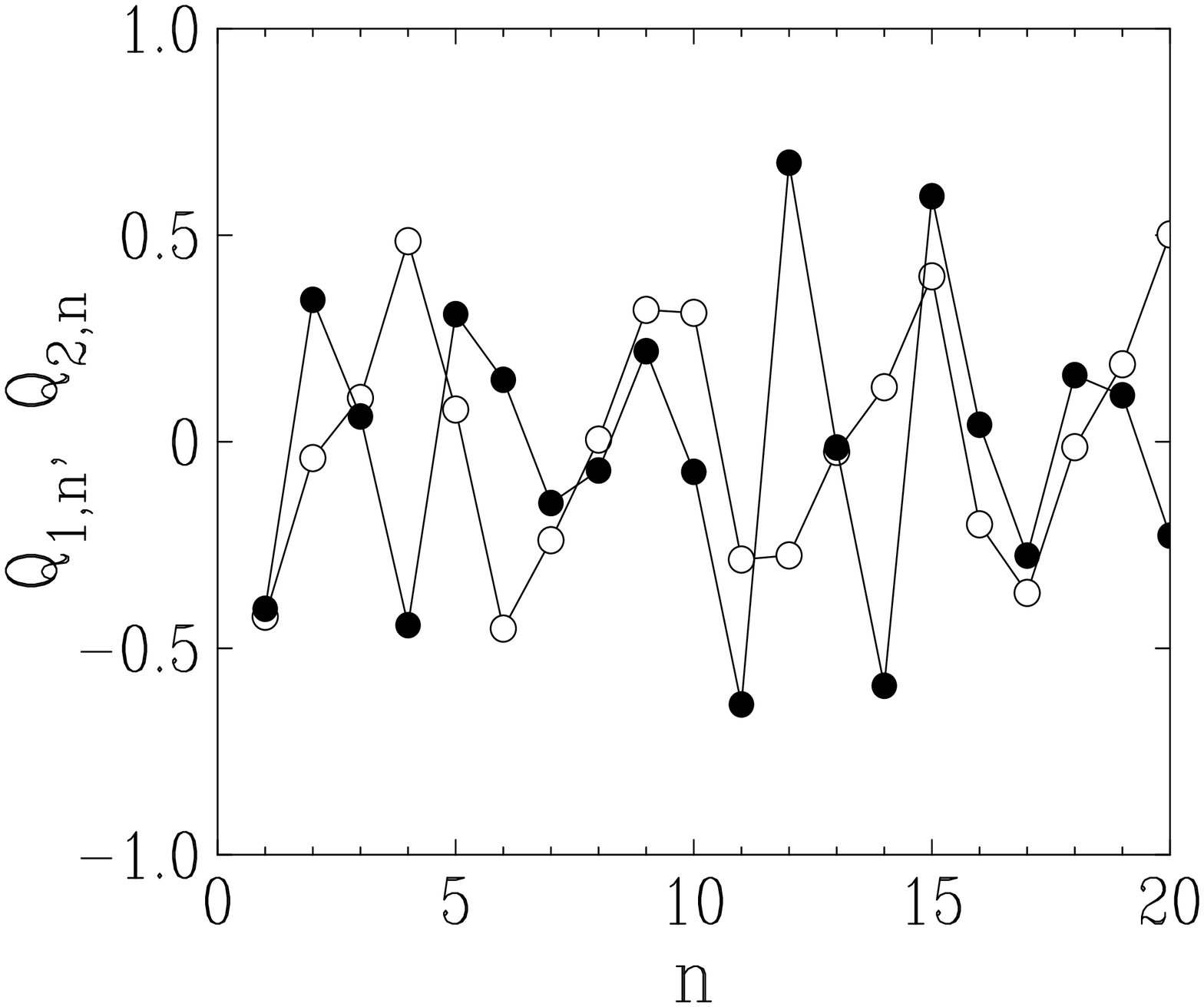}
\caption{Plots of the current amplitudes $Q_{1,n}$ (full symbols)
and $Q_{2,n}$ (empty symbols)
against level number $n$, for $\Phi_1=BL_1^2/(4\pi)=\pi/3$.
Left: commensurate case $L_2=2L_1$, i.e., $\o=1/3$.
Right: incommensurate case $\o=1/\tau^2$.}
\label{figlevels}
\end{center}
\end{figure}

\begin{figure}[!ht]
\begin{center}
\includegraphics[angle=90,width=.4\linewidth]{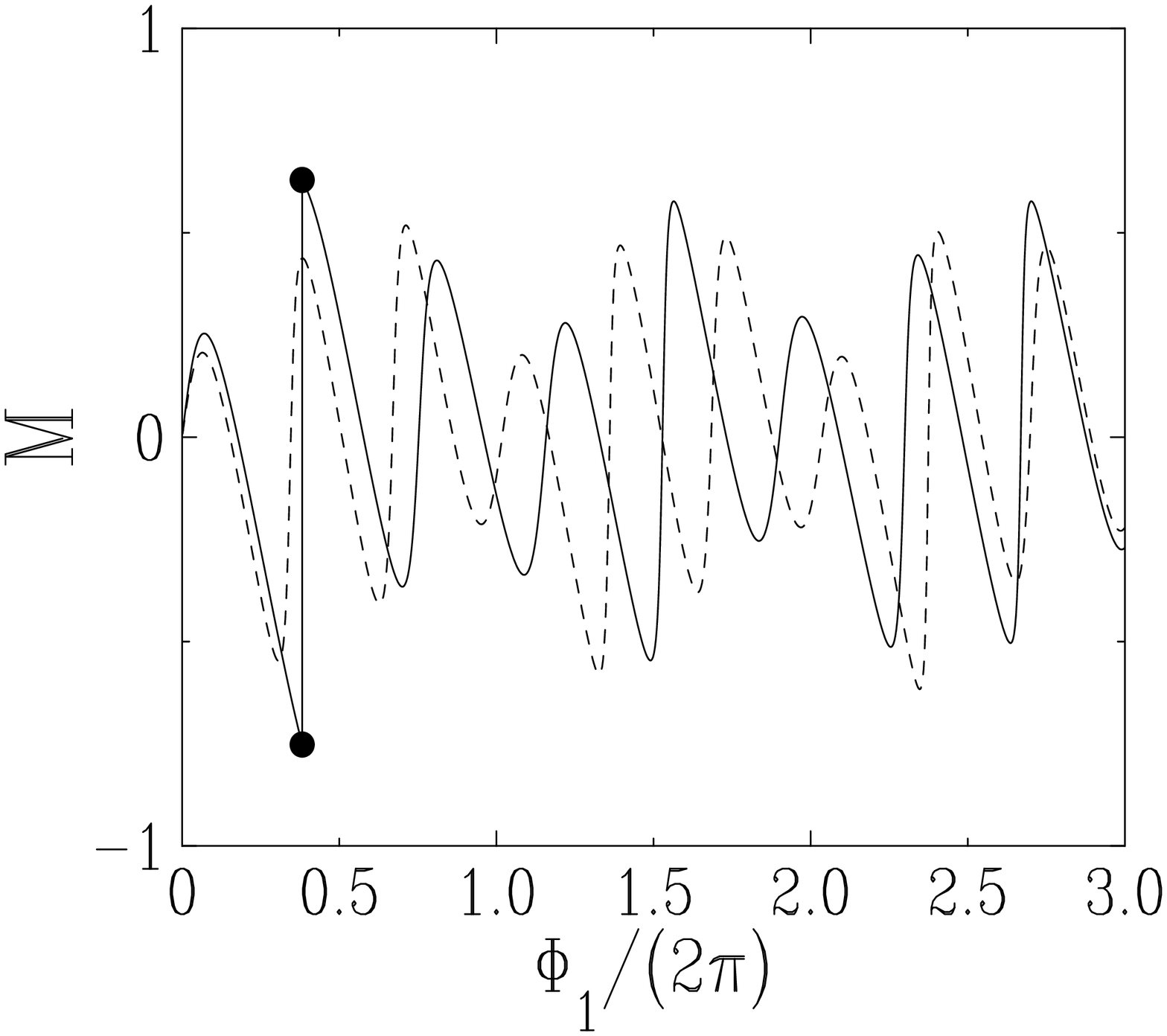}
{\hskip 10pt}
\includegraphics[angle=90,width=.4\linewidth]{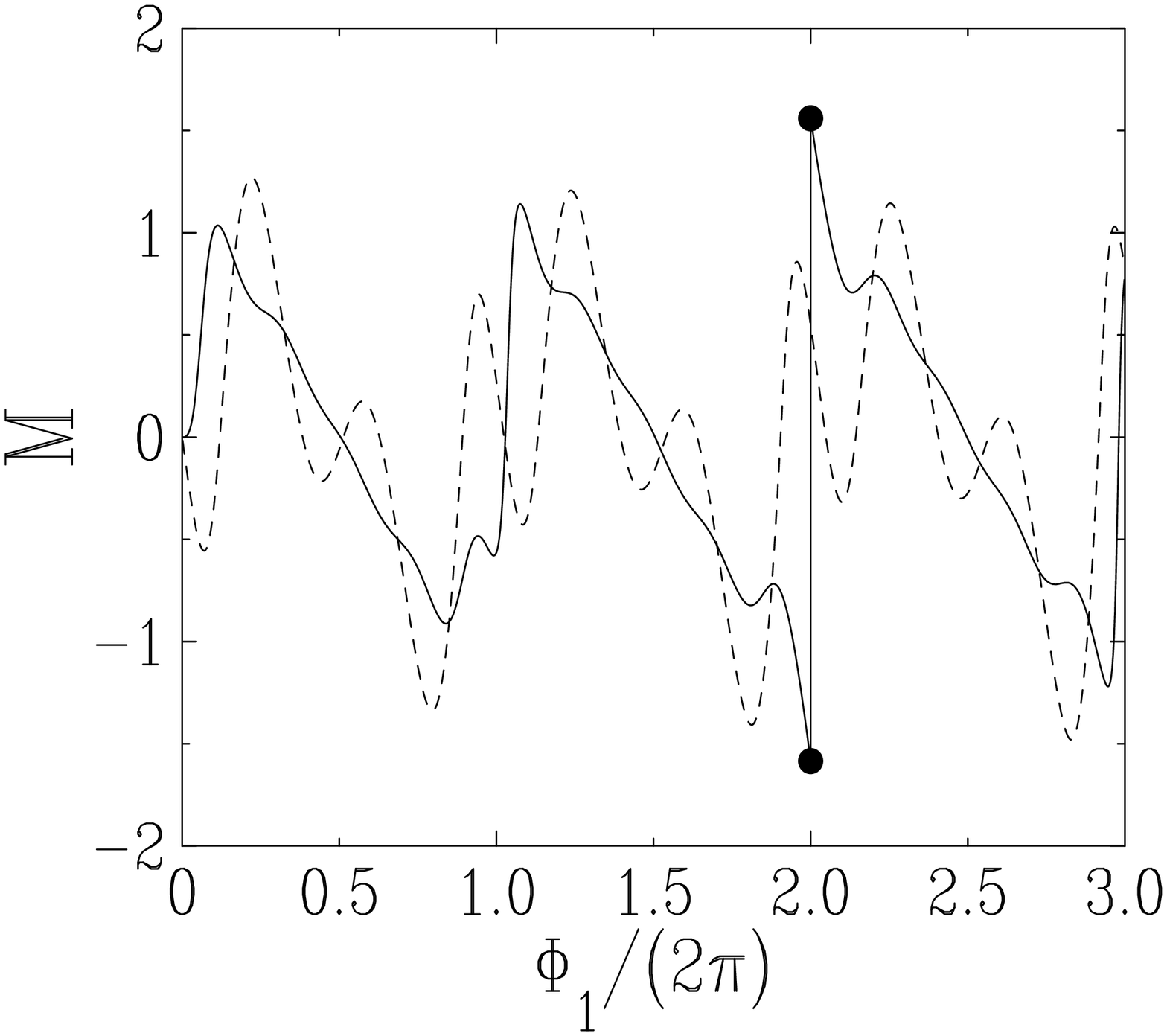}
\caption{Plots of the magnetization $M$ against $\Phi_1/(2\pi)=BL_1^2/(8\pi^2)$,
for $N=3$ (left) and $N=10$ (right).
Full lines: $\o=1/\tau^2$
(jumps due to twofold degeneracies are shown by symbols).
Dashed lines: $\o=1/\e$.}
\label{figm}
\end{center}
\end{figure}

Let us close up this section with some numerical illustrations
of our results in the case of circular rings.
The observables (persistent currents and magnetization)
are given in terms of the hull functions by~(\ref{qqp}),
and (\ref{im}), (\ref{total}).
Figure~\ref{figlevels} shows plots of the current amplitudes
$Q_{1,n}$ and $Q_{2,n}$ of individual levels against level number $n$,
for a system of two circular rings with $\Phi_1=BL_1^2/(4\pi)=\pi/3$,
in a typical commensurate case (left): $L_2=2L_1$, i.e., $\o=1/3$,
and in a typical incommensurate case (right): $\o=1/\tau^2$.
The period $2m=6$ predicted in Section~\ref{com}
is clearly observed in the commensurate case.
Figure~\ref{figm} shows plots of the magnetization
against $\Phi_1/(2\pi)=BL_1^2/(8\pi^2)$, for $N=3$ (left) and $N=10$ (right),
for two incommensurate cases corresponding to the nearby irrationals
$\o=1/\tau^2\approx0.381\,966$ and $\o=1/\e\approx0.367\,879$.
Twofold degeneracies are observed in the first case,
in agreement with~(\ref{wquadra}), as $\o=1/\tau^2$ obeys $\o^2-3\o+1=0$.

\section{The r\^ole of spin-orbit interaction}
\label{so}

So far spin did not play any role in our discussion.
We now investigate the case that,
besides the Abelian U(1) magnetic fluxes $\Phi_1$ and $\Phi_2$,
there are also non-Abelian SU(2) fluxes $\Psi_1$ and $\Psi_2$.

For the sake of consistency, let us recall some basic notions
pertaining to the physics of SU(2) fluxes.
Such fluxes arise as a result of spin-orbit interaction.
Unlike U(1) fluxes, SU(2) fluxes are invariant under time reversal.
While the U(1) flux leading to the Aharonov-Bohm effect
is realized by threading a ring with a magnetic field,
the SU(2) flux leading to the Aharonov-Casher effect is generated by piercing
a ring with a line of charge.
More precisely,
if a system of electrons is confined to a plane
and subject to an electric field generated by a
straight perpendicular charged wire
with constant charge $\lambda$ per unit length,
we have the SU(2) analogue of the Aharonov-Bohm effect.
The starting point of the analysis is the Pauli Hamiltonian.
In the presence of a vector potential $\ve A$ and of an electric field~$\ve E$,
within the approximate $\mathrm{U(1)}\otimes\mathrm{SU(2)}$ symmetry
of the non-relativistic Schr\"odinger equation~\cite{froh},
this Hamiltonian reads, in dimensionful form,
\beq
H_{\mathrm Pauli}=\frac{1}{2m}
\left(\ve p+\frac{e}{c}{\ve A}+\gamma\hbar\ve E\times\ve\si\right)^2,
\label{Paulieq}
\eeq
where $\gamma=e/(4mc^2)$.
The third term on the right-hand side of~(\ref{Paulieq})
is responsible for the spin-orbit interaction.
In the case of a circular ring of radius $R$
pierced by a charged wire through its center,
we have $\ve E=2\lambda\ve r/R^2$,
whereas the curvilinear abscissa and the circumference read
$s=R\theta$ and $L=2\pi R$.
In this geometry, the U(1) and SU(2) potentials appearing in~(\ref{Paulieq})
can be eliminated {\it locally} by the respective gauge transformations
\beq
\matrix{
\ds g_{\mathrm U(1)}
=\exp\left(-\frac{\i e}{\hbar c}\int_0^s\ve A\cdot\d\ve r\right)
=\exp\left(-\frac{\i\Phi\theta}{2\pi}\right),\hfill\cr
\ds g_{\mathrm SU(2)}
=\exp\left(-\i\gamma\int_0^s(\ve E\times\ve\si)\cdot\d\ve r\right)
=\exp\left(-\frac{\i\Psi\,\ve\hatn\cdot\ve\si\,\theta}{2\pi}\right),}
\label{gaugetransformation}
\eeq
where the integrations are carried out along the ring.
In the above expression for $g_{\mathrm SU(2)}$,
the integral need not be path-ordered.
We have indeed
$(\ve E\times\ve\si)\cdot\d\ve r
=-(\ve E\times\d\ve r)\cdot\ve\si
=-(2\lambda/R^2)(\ve r\times\d\ve r)\cdot\ve\si
=-2\lambda\,\ve\hatn\cdot\ve\si\,\d\theta$,
where $\ve\hatn$ is the (properly oriented)
unit vector perpendicular to the plane of the ring.
As a result, the dimensionless SU(2) flux reads $\Psi=-4\pi\gamma\lambda$.
The above simplifying property is specific of the case
where the electric field lies in the plane in which the electrons are confined.
It has also been used in a spherical geometry,
in deriving a tight-binding version of the $\ve L\cdot\ve S$ spin-orbit
interaction~\cite{AL2}.
The SU(2) flux appears as a `pure gauge',
in the sense that it only affects the phase of the wavefunction.
However, the U(1) and SU(2) potentials cannot be eliminated {\it globally}.
These potentials rather give rise to non-integrable phase factors,
$\e^{\i\Phi}$ for U(1) and $\e^{\i\Psi\,\ve\hatn\cdot\ve\si}$ for SU(2).
Observables are therefore $2\pi$-periodic both in $\Phi$ and in $\Psi$.
The upshot of the above discussion is that, in the ring geometry,
the Pauli Hamiltonian can be replaced by a simpler one, namely
\beq
H=\left(p-a-b\,\ve\hatn\cdot\ve\si\right)^2,
\label{Paulieqring}
\eeq
with the notations $p=-\i\d/\d s$, $a=\Phi/L$ and $b=\Psi/L$.
For a single ring in the $(x,y)$ plane,
we have $\ve\hatn=\ve\hatz$, so that $s_z$ is conserved.
The problem then reduces to that of two independent Aharonov-Bohm
systems of polarized electrons, one with flux $\Phi+\Psi$ and the second with
flux $\Phi-\Psi$.

\begin{figure}[!ht]
\begin{center}
\includegraphics[angle=0,width=.4\linewidth]{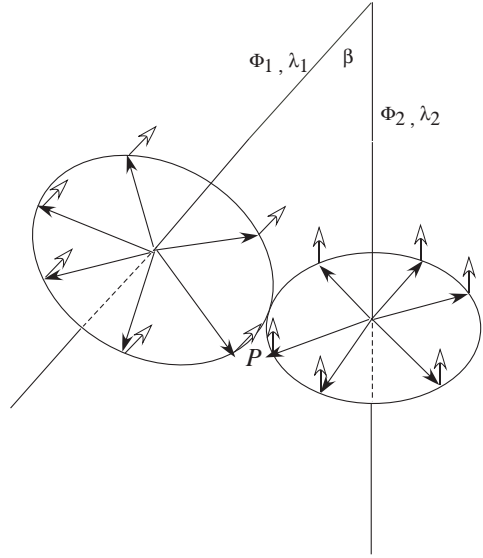}
\caption{Gedankenexperimental realization of the two rings
with U(1) and SU(2) fluxes such that spin is not conserved.
Solid lines
represent magnetic fluxes $\Phi_1$ and $\Phi_2$ as well as charged wires
with respective longitudinal charge densities $\lambda_1$ and $\lambda_2$.
Full arrows mark the direction of the electric fields,
whereas empty arrows
indicate the directions of the spins (more precisely, ${\hat {\bf n}}_{1,2}$).
This is the
special case corresponding to the Hamiltonian (\ref{HSU2}) where the two axes
${\hat {\bf n}}_{1,2}$
lie in the same $(x,z)$ plane.}
\label{SU2RINGS}
\end{center}
\end{figure}

A somewhat less simple manifestation of the AC effect in terms of SU(2) fluxes
occurs when $s_z$ is not conserved.
This can be realized (for example) by considering two rings in different planes
respectively subject to perpendicular magnetic fields and pierced by
perpendicular lines of charges with uniform densities $\lambda_1$ and
$\lambda_2$.
Such a Gedankenexperiment is schematically displayed in
Figure~\ref{SU2RINGS}.
Thus, while in the U(1) two-ring problem the geometric configuration of the two
rings (in particular their relative orientation)
did not play an important r\^ole, the sample's geometry becomes of central
importance when SU(2) fluxes are considered.
This situation allows one to study the influence
of SU(2) fluxes on the U(1) magnetization~\cite{Meir}.
The main messages of the analysis given below are as follows.
(i) Even when $\ve\hatn_1$ is antiparallel to $\ve\hatn_2$,
the non-conservation of $s_z$ requires {\it both} U(1) fluxes
to be non-trivial, i.e., neither integer nor half-integer.
This is a novel situation where the energy levels are sensitive to a
combination of AB and AC effects.
(ii) Even though $\Psi$ enters as a pure gauge, spin-orbit interaction
might change the sign of the magnetization,
so that the system switches between a paramagnetic and a diamagnetic response.
(iii) In complete analogy with the U(1) case,
where the orbital magnetization is related to the derivative
of the ground-state energy with respect to $\Phi$,
it is natural to define an `SU(2) magnetization'
related to the derivative of the ground-state energy with respect to $\Psi$.
This magnetization is another equilibrium property.
In each ring and for a given level $n$,
it is proportional to the expectation value of the
commutator $\{\hat v,\ve\hatn\cdot\ve\si\}$,
where ${\hat v}$ is the velocity operator.
Thus, it can in principle be measured, despite the fact that the SU(2) (spin)
current is not conserved while the U(1) (charge) current is.

The one-electron Hamiltonian of our two-ring system now reads
\beq
\H=\left(p_1-a_1-b_1\,\ve\hatn_1\cdot\ve\si\right)^2
+\left(p_2-a_2-b_2\,\ve\hatn_2\cdot\ve\si\right)^2,
\label{HSU2}
\eeq
where $p_1$ and $p_2$ are the differential operators of~(\ref{pdef}),
$\ve\hatn_1$ and $\ve\hatn_2$ are two arbitrary unit vectors,
$\ve\si$ is the vector of Pauli matrices,
whereas the U(1) vector potentials $a_1$ and $a_2$
and their SU(2) analogues $b_1$ and $b_2$
are related to the corresponding fluxes as follows:
\beq
a_1=\frac{\Phi_1}{L_1},\quad a_2=\frac{\Phi_2}{L_2},\quad
b_1=\frac{\Psi_1}{L_1},\quad b_2=\frac{\Psi_2}{L_2},
\eeq
The problem mostly depends on the angle $\beta$
between both directions $\ve\hatn_1$ and $\ve\hatn_2$ in spin space,
such that $\ve\hatn_1\cdot\ve\hatn_2=\cos\beta$.
For convenience we choose axes such that
$\ve\hatn_1=(0,0,1)$ is along the $z$-axis
whereas $\ve\hatn_2=(\sin\beta,0,\cos\beta)$ is in the $(x,z)$ plane.

A state $\vert\psi\rangle$ is described by a pair of wavefunctions
$\{\ve\psi^\un(s_1),\;\ve\psi^\de(s_2)\}$,
each of them being a two-component spinor.
Separating spin and orbital degrees of freedom,
we are led to look for an eigenstate of $\H$ in the form
\beqa
\ve\psi^\un(s_1)
&=&\pmatrix{1\cr0}\e^{\i(a_1+b_1)s_1}(A_1\,\e^{\i qs_1}+B_1\,\e^{-\i qs_1})\cr
&+&\pmatrix{0\cr1}\e^{\i(a_1-b_1)s_1}(C_1\,\e^{\i qs_1}+D_1\,\e^{-\i qs_1}),\cr
\ve\psi^\de(s_2)
&=&\pmatrix{\cos\frac{\beta}{2}\cr\sin\frac{\beta}{2}}\e^{\i(a_2+b_2)s_2}
(A_2\,\e^{\i qs_2}+B_2\,\e^{-\i qs_2})\cr
&+&\pmatrix{-\sin\frac{\beta}{2}\cr\cos\frac{\beta}{2}}\e^{\i(a_2-b_2)s_2}
(C_2\,\e^{\i qs_2}+D_2\,\e^{-\i qs_2}).
\eeqa
Along the lines of Section~\ref{secham},
the continuity conditions generalizing~(\ref{pcont})
allow one to express the eight amplitudes $A_1,\dots,D_2$
in terms of the two components of~$\ve\psi(\PT)$.
The current conservation conditions generalizing~(\ref{pcur})
then yield the characteristic equation $D\so(q)=0$, with
\beqa
D\so(q)
&=&\Bigl(\sin q(L_1+L_2)
-\cos(\Phi_2+\Psi_2)\sin qL_1-\cos(\Phi_1+\Psi_1)\sin qL_2\Bigr)
\nonumber\\
&\times&\Bigl(\sin q(L_1+L_2)
-\cos(\Phi_2-\Psi_2)\sin qL_1-\cos(\Phi_1-\Psi_1)\sin qL_2\Bigr)
\nonumber\\
&+&4\,\sin qL_1\,\sin qL_2\,\sin\Phi_1\,\sin\Phi_2\,\sin\Psi_1\,\sin\Psi_2
\,\sin^2\frac{\beta}{2}.
\eeqa

The first two lines of this expression are identical to the scalar
characteristic equation~(\ref{dres}), up to the replacement of the magnetic
fluxes by the sums and differences of their Abelian and non-Abelian parts:
$\Phi_1\to\Phi_1\pm\Psi_1$, $\Phi_2\to\Phi_2\pm\Psi_2$.
Each factor therefore describes a scalar problem
in an effective Abelian flux.
The third line provides the coupling between both spin components.
In order for this coupling to be non-zero,
the four fluxes need to be simultaneously non-trivial,
i.e., not equal to 0 or $\pi\mod{2\pi}$,
and the angle $\beta$ not equal to $0\mod{2\pi}$.

In the case of two equal rings
($L_1=L_2=L$, $\Phi_1=\Phi_2=\Phi$, $\Psi_1=\Psi_2=\Psi$),
the spectrum consists of an alternation of groups of two degenerate idle states,
such that $\sin qL=0$, i.e., $qL=p\pi$,
and of groups of two non-degenerate active states, such that
\beq
\cos qL=\cos\theta_\pm=\cos\Phi\cos\Psi\pm\sin\Phi\sin\Psi\cos\frac{\beta}{2},
\eeq
Choosing for definiteness $0\le\beta\le\pi$,
we have $0\le\theta_+\le\theta_-\le\pi$.
For $N=2$ electrons, only the first two active states are occupied.
The magnetization reads
\beq
M=-\frac{1}{2\pi}\left(\theta_+\,\frac{\dpar\theta_+}{\dpar\Phi}
+\theta_-\,\frac{\dpar\theta_-}{\dpar\Phi}\right).
\eeq

\begin{figure}[!ht]
\begin{center}
\includegraphics[angle=90,width=.4\linewidth]{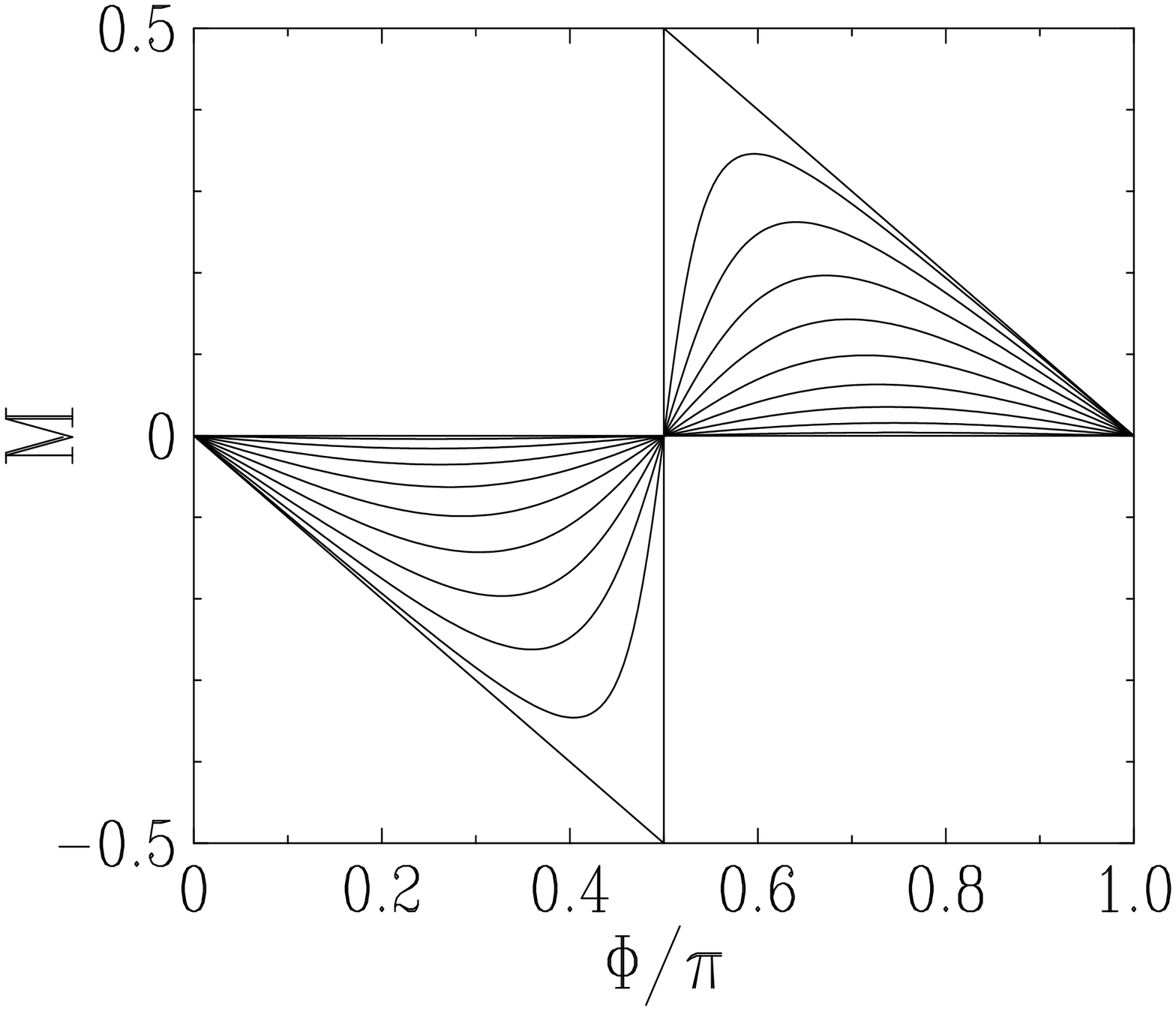}
{\hskip 10pt}
\includegraphics[angle=90,width=.4\linewidth]{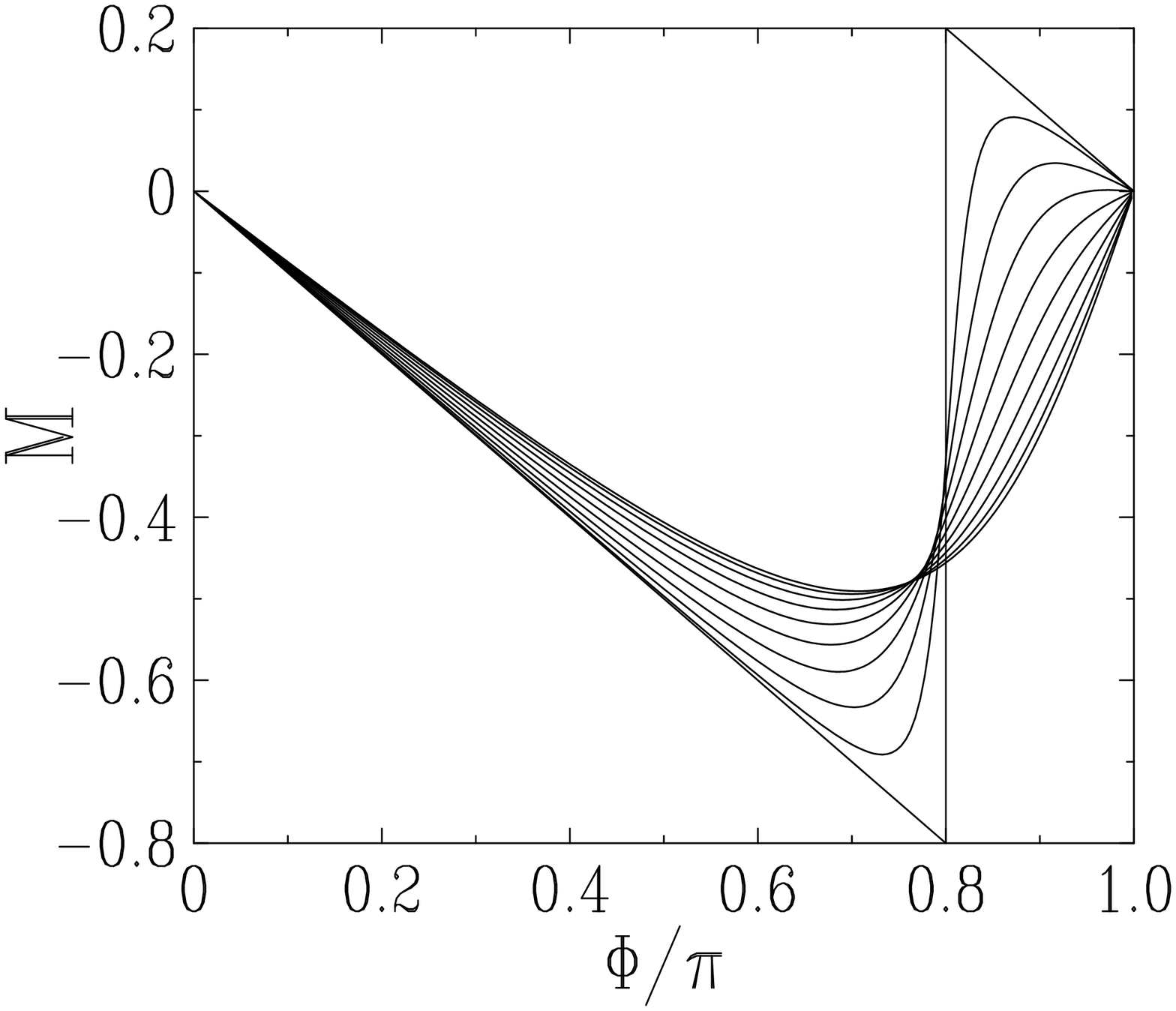}
\caption{Plots of the magnetization $M$ of two electrons on two equal rings,
against $\Phi/\pi$, for $\Psi=\pi/2$ (left) and $\Psi=\pi/5$ (right).
In both cases, $\beta=k\pi/10$, with $k=0,\dots,10$,
bottom to top in the left part of the curves.}
\label{figso}
\end{center}
\end{figure}

\begin{figure}[!ht]
\begin{center}
\includegraphics[angle=90,width=.4\linewidth]{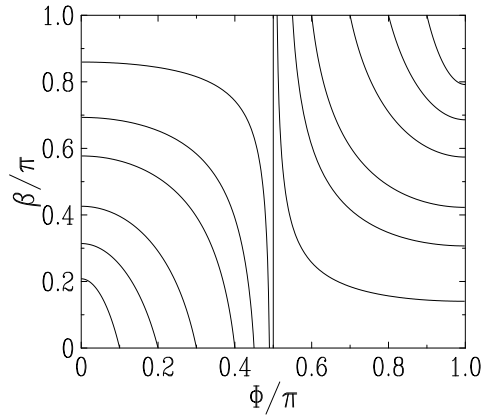}
\caption{Plot of the locus in the $(\Phi,\beta)$ plane
where the magnetization $M$ of two electrons on two equal rings vanishes,
at fixed $\Psi$.
From top right to bottom left, $\Psi/\pi=0.1$, 0.2, 0.3, 0.4, 0.45, 0.49,
0.5 (vertical line), 0.51, 0.55, 0.6, 0.7, 0.8 and 0.9.}
\label{figzero}
\end{center}
\end{figure}

Let us consider the dependence of $M$ on the Abelian flux $\Phi$
in the range $0\le\Phi\le\pi$ and for $0\le\Psi\le\pi$.
For $\beta=0$, we obtain
\beq
M=\left\{\matrix{
-\Phi/\pi\hfill&\hbox{for}&0\le\Phi<\pi-\Psi,\cr
(\pi-\Phi)/\pi&\hbox{for}&\pi-\Psi<\Phi\le\pi.\hfill
}\right.%}
\eeq
This discontinuous jump in the magnetization
is rounded for non-zero values of $\beta$, as shown in Figure~\ref{figso}.
This figure also illustrates a remarkable feature of such a simple system
of two electrons on two equal rings,
namely that the magnetization changes sign as a function of parameters.
For a small magnetic flux $\Phi$, we have
\beq
M=-\frac{1}{\pi}
\left(\cos^2\frac{\beta}{2}+\sin^2\frac{\beta}{2}\,\Psi\cot\Psi\right)\Phi
+\cdots
\eeq
The expression in the parentheses vanishes for
\beq
\tan^2\frac{\beta}{2}=-\frac{\tan\Psi}{\Psi},
\eeq
provided $\Psi>\pi/2$.
More generally, the magnetization vanishes along a $\Psi$-dependent curve
in the $(\Phi,\beta)$ plane, shown in Figure~\ref{figzero}.
At fixed $\Psi$,
the magnetization is positive above and to the right of that curve,
whereas it is negative below and to the left of that curve.

\section{Discussion}
\label{Sec:Discussion}
The many differences between the single and the double-ring geometry,
which have been underlined throughout this work,
testify the importance of topology in mesoscopic physics.
From a local viewpoint,
both systems are governed by the same one-dimensional Schr\"odinger equation.
They only differ in their global topological structure,
measured by their genus (number of holes).

Another example underlining the key r\^ole of topology
in the context of persistent currents has been
studied by a collaboration involving one of us~\cite{YAMobius}.
The importance of the sample's topology in mesoscopic physics
has already been stressed by Schmeltzer~\cite{Schmeltzer},
who studied the persistent current of spinless fermions
in a double-ring system, using the formalism of Dirac constraints.
In fact, Schmeltzer's work was the main motivation for the present research.
Apart from underlining the importance of topology,
the work~\cite{Schmeltzer} however focussed on different aspects.
In particular, the r\^ole of spin and the AC effect were not tackled.

We have not tackled the rich physics that should emerge
when interaction effects are taken into account.
First of all, there are interesting charging effects,
suggested and discussed for the single-ring geometry
in~\cite{Buttiker1,Buttiker2}.
Second, interactions in one dimension turn the system to a non-Fermi liquid.
While this problem has been studied for a single ring~\cite{Loss},
persistent currents in interacting quantum rings
have recently been studied in~\cite{Peeters}.
It can however be anticipated that the detailed analysis presented here
for the non-interacting system
will turn its place to a complicated and somewhat intractable formalism,
so that e.g.~the fine effects of length and flux commensurability
will be absent.

The two-ring geometry provides a playground for testing
fundamental aspects of quantum mechanics,
such as the occurrence of interlaced AB and AC effects.
At the same time, a two-ring system is certainly within the reach of fabrication
(see Figure~1 in~\cite{Russians}),
so that the present study is also rooted into the real world.
At the same time, the difficulties of controlling the external electric field
in semiconductor he\-te\-ro\-structures hosting a two-dimensional electron gas,
as well as their enhancement by internal fields of ion cores and discontinuities
(eventually manifest as spin-orbit couplings)
have recently been summarized in a comprehensive review article~\cite{Fabian}.
Thus, while the bare Hamiltonian~(\ref{HSU2}) of course applies
for any field whatsoever,
a realization of the two-ring geometry as in Figure~\ref{SU2RINGS}
is still to be considered as a Gedankenexperiment.
Nevertheless, the main conclusion of our analysis is fundamental
and should pass any experimental test, namely, an AC effect which is
a pure gauge (periodic in the SU(2) phase) and does not conserve $s_z$
is realizable only if the AB effect is present.

Finally, although the present work has focussed onto zero-temperature
equilibrium properties,
it can be expected that transport properties will reveal
a similar richness of behavior.

\ack

We would like to thank D~Schmeltzer for having brought this subject
to our attention and communicated to us preliminary versions
of his article~\cite{Schmeltzer}.
Discussions with G~Montambaux and X~Waintal are also warmly acknowledged.

This work was initiated during a visit of YA
supported by RTRA -- Triangle de la Physique
to the Institut de Physique Th\'eorique (CEA, Saclay)
and to the Laboratoire de Physique des Solides (Universit\'e Paris-Sud, Orsay).

\appendix

\section{One single ring: a reminder}

In this Appendix we give a brief reminder of the well-known
situation of a single clean ring.
The ring with length $L$ and area $A$ is threaded by a flux $\Phi=BA$.
It may assume an arbitrary planar shape.
With the same conventions as in the body of this work,
parametrizing a point of the wire
by its curvilinear abscissa $s$ in the range $0\le s\le L$,
the one-body Hamiltonian reads
\beq
\H=(p-a)^2,
\eeq
with $p=-\i\,\d/\d s$ and $a=\Phi/L$.

It is worth mentioning the analogy with Bloch theory,
already noticed in~\cite{BIL}.
Starting from the Schr\"odinger equation $\H\psi=E\psi$
with $\psi(s+L)=\psi(s)$,
a gauge transformation $\psi(s)=\e^{\i as}\eta(s)$
leads to the following equation and boundary condition for $\eta(s)$:
\beq
p^2\eta(s)=E\eta(s),\quad\eta(s+L)=\e^{\i\Phi}\eta(s).
\label{Eq_gauge}
\eeq
This is exactly the equation for an electron in a one-dimensional lattice
potential of period $L$ and Bloch wavenumber $K$ such that $KL=\Phi$,
i.e., $K=a=\Phi/L$.
One immediate consequence is that the energy is periodic in $\Phi$
with period $2\pi$.
This results also holds when the ring is not clean,
since the full (disordered) potential
can still be viewed as a periodic potential of period~$L$.

The eigenstates of $\H$ are given by the wavefunctions
\beq
\psi(s)\sim\e^{\i(q+a)s}.
\eeq
The periodicity of $\psi(s)$ in $s$ with period $L$
yields the quantization condition $(q+a)L=2\pi k$, hence
\beq
q_k=\frac{2\pi k-\Phi}{L},
\eeq
where $k=0,\pm1,\dots$
The wavefunctions $\psi_k(s)\sim\exp(2\pi\i ks)$
therefore do not depend on the magnetic flux $\Phi$.

The contributions $I_k$ and $M_k$ of the eigenstate number $k$
to the persistent current and the magnetization read
\beq
I_k=-\frac{\dpar E_k}{\dpar\Phi}=\frac{2q_k}{L},\quad
M_k=-\frac{\dpar E_k}{\dpar B}=AI_k.
\label{imres}
\eeq
With the notation~(\ref{im}), the above result for $I_k$ simply reads $Q_k=1$.

For a zero-temperature system with $N$ electrons,
and for $0\le\Phi\le\pi$, we have the following results.

\begin{itemize}

\item
For $N=2p-1$ odd, the occupied states are $k=-p+1,\dots,p-1$.
We obtain
\beq
E=E_0+\frac{N}{L^2}\,\Phi^2,\quad M=-\frac{2A}{L^2}\,N\Phi.
\label{sodd}
\eeq

\item
For $N=2p$ even, the occupied states are $k=-p+1,\dots,p$.
We obtain
\beq
E=E_0+\frac{N}{L^2}\,(\pi-\Phi)^2,\quad M=\frac{2A}{L^2}\,N(\pi-\Phi).
\label{seven}
\eeq

\end{itemize}

In both cases the minimum energy reads $E_0=N(N^2-1)\pi^2/(3L)$.
For a circular ring with radius $R$,
we have $L=2\pi R$ and $A=\pi R^2$, so that $2A/L^2=1/(2\pi)$.

\section{Three coupled rings}

\begin{figure}[!ht]
\begin{center}
\includegraphics[angle=90,width=.4\linewidth]{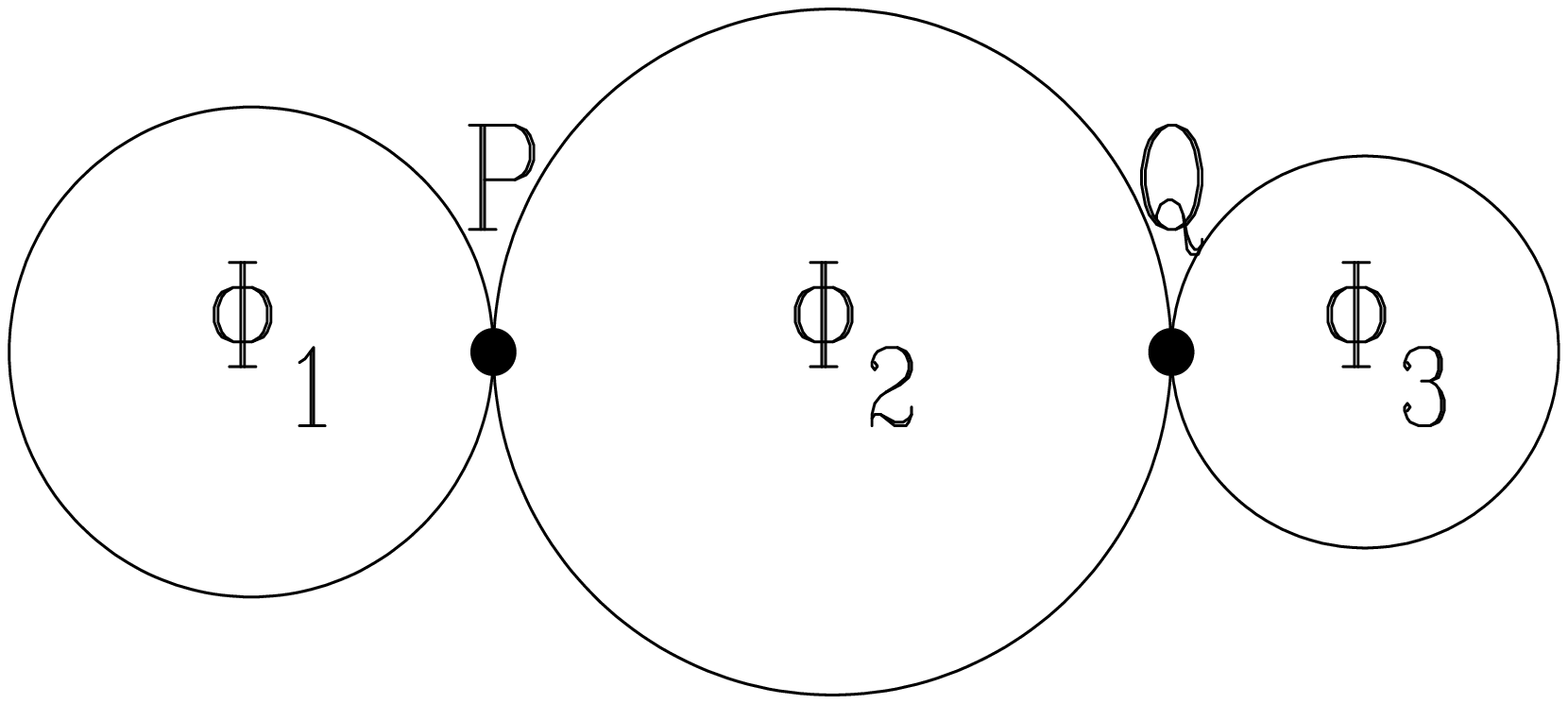}
\caption{The sample considered in Appendix~B is made of three unequal rings
touching at two contact points P and Q.}
\label{figthree}
\end{center}
\end{figure}

In this Appendix we show how the present investigation
can be extended to more complex geometries.
We consider for definiteness the case of a sample
made of three unequal rings touching at two contact points P and Q,
as shown in Figure~\ref{figthree}.
The line PQ is assumed to be an axis of symmetry of the sample.
The left, middle and right rings have respective lengths
$L_1$, $L_2$ and $L_3$ and areas $A_1$, $A_2$ and $A_3$.
They are therefore threaded by magnetic fluxes
$\Phi_1=BA_1$, $\Phi_2=BA_2$ and $\Phi_3=BA_3$.

The one-electron Hamiltonian of the system reads
\beq
\H=(p_1-a_1)^2+(p_2-a_2)^2+(p_3-a_3)^2,
\eeq
with $p_1=-\i\,\d/\d s_1$, $p_2=-\i\,\d/\d s_2$, $p_3=-\i\,\d/\d s_3$,
$a_1=\Phi_1/L_1$, $a_2=\Phi_2/L_2$, $a_3=\Phi_3/L_3$.

A state is now described by three wavefunctions, one living on each ring:
$\{\psi^\un(s_1),\;\psi^\de(s_2),\;\psi^\trois(s_3)\}$.
These wavefunctions obey two continuity conditions of the form~(\ref{pcont})
(one at P and one at Q)
and two current conservation conditions of the form~(\ref{pcur}).
Along the lines of Section~\ref{secham},
we obtain the characteristic function
\beqa
D_3(q)
&=&\cos q(L_1+L_2+L_3)\cr
&-&\cos\Phi_3\cos q(L_1+L_2)-\cos\Phi_1\cos q(L_2+L_3)\cr
&+&\cos\Phi_3\cos qL_1+\cos\Phi_1\cos\Phi_3\cos qL_2+\cos\Phi_1\cos qL_3\cr
&+&\cos\Phi_2\sin qL_1\sin qL_3-\cos qL_1\cos qL_3-\cos\Phi_1\cos\Phi_3.
\label{d3}
\eeqa

Many of the outcomes of this paper can be extended to the present situation,
although expressions become very cumbersome.
The energy eigenvalues $E_n=q_n^2$ can be parametrized as
\beq
q_n=\frac{n\pi+g_n}{L_1+L_2+L_3}\quad(n=1,2,\dots),
\eeq
where the modulation $g_n$ obeys the bounds $-\pi\le g_n\le2\pi$.
In the absence of magnetic fluxes,
the factorized form~(\ref{dfactor}) of the characteristic function
generalizes to
\beq
D_3(q)=8\,\sin\frac{q(L_1+L_2+L_3)}{2}
\,\sin\frac{qL_1}{2}\,\sin\frac{qL_2}{2}\,\sin\frac{qL_3}{2}.
\eeq
The spectrum consists of four sectors.
The corresponding states can be respectively referred to as
{\it trilateral, left, central} and~{\it right}.

\end{document}